\documentclass[desactivate]{aa}  

\usepackage{rotating}
\usepackage{newtxtext,newtxmath}
\usepackage{graphicx}	
\usepackage{amssymb}	
\usepackage{tabularx}
\usepackage{natbib}
\usepackage{booktabs}
\usepackage{multirow}
\usepackage{comment}
\usepackage{lscape}
\usepackage{color}
\usepackage[utf8]{inputenc}
\usepackage{hyperref}

\usepackage[T1]{fontenc}
\def\cstat{{\sc C-stat}}
\def\dcstat{$\Delta$\cstat}

\begin{document}

   \title{Probing for relativistic outflows in the Ultraluminous X-ray pulsar RX J0209.6-7427}

   \author{F. Barra\inst{1,2,3}\fnmsep\thanks{ francesco.barra@inaf.it}, C. Pinto\inst{1}, T. di Salvo\inst{2}, S. Caserta\inst{2,1} \and P. Kosec\inst{3}
          }
   \institute{INAF - IASF Palermo, via Ugo La Malfa 153, I-90146 Palermo, Italy 
\and Universit\`a degli Studi di Palermo, Dipartimento di Fisica e Chimica, via Archirafi 36, I-90123 Palermo, Italy 
\and Center for Astrophysics | Harvard  \& Smithsonian, 60 Garden Street, Cambridge, MA 02138, USA
\\
              }
   \date{XXXX; accepted YYYY}

% \abstract{}{}{}{}{} 
% 5 {} token are mandatory
 
  \abstract
  % context heading (optional)
  % {} leave it empty if necessary  
   {The launching of winds from accretion discs arises naturally from the accretion process onto compact objects. Such outflows have been observed in Galactic X-ray binaries (XRBs) and play an even more significant role in extragalactic systems, namely Ultra-luminous X-ray sources (ULXs), which are characterised by powerful, relativistic winds. At large distances, however, their detection rate remains limited, and the physical processes driving these winds are challenging to investigate in detail. Galactic neutron stars accreting matter from Be-star companions provide a valuable opportunity to improve our understanding of these outflows.}
  % aims heading (mandatory)
   {RX J0209.6-7427 is a Be-XRB, powered by a neutron star with a spin period of 9.3 seconds, that underwent an outburst in 2019, exceeding a bolometric luminosity of $10^{39} \ \rm erg \ s^{-1}$ at its peak. At such extreme luminosities, relativistic outflows are expected, as observed in ULXs. Owing to its relatively low distance and the availability of a high-resolution XMM-\textit{Newton} observation, this source represents a strong candidate for investigating and detecting outflows associated with super-Eddington accretion.}
  % methods heading (mandatory)
   {The RGS spectrum yields 105,000 counts, a factor of two higher than any other deep ULX RGS spectrum. Both the RGS and EPIC-pn datasets were modelled using a phenomenological continuum that accounts for emission from the inner and outer regions of the accretion disc as well as from the accretion column, following standard practice in ULX studies. The emission and absorption features were subsequently fitted with the optically-thin plasma models available in {\scriptsize{SPEX}}, performing an extensive exploration of the temperature–velocity parameter space in order to probe Doppler shifts.}
  % results heading (mandatory)
   {The modelling of the spectral features in emission suggests the presence of a multiphase photoionised plasma approximately at rest with respect to the source. The absorption features can be reproduced either by blueshifted photoionised absorbers moving at approximately $0.1c$ and $0.2c$, the latter being more tentative, or by an alternative collisionally ionised absorption model, which provides a comparably acceptable fit. The presence of ultrafast outflows therefore remains a plausible, but non-unique, interpretation of the data. If associated with outflows, the inferred column densities appear lower than those reported for some persistent ULXs, potentially reflecting differences in the accretion-flow properties or the shorter super-Eddington phase of RX J0209.6-7427. }
  % conclusions heading (optional), leave it empty if necessary 
   {}
\authorrunning{F. Barra et al. }
\titlerunning{Probing for relativistic outflows in the Ultraluminous X-ray pulsar RX J0209.6-7427}

   \maketitle

%-------------------------------------------------------------------

\section{Introduction}
Ultraluminous X-ray sources (ULXs) are mainly extragalactic, off-nuclear sources characterised by an X-ray luminosity $\ge 10^{39} \ \rm erg/s$ and higher than the Eddington limit for an accreting 10 M\textsubscript{\(\odot\)} black hole (BH), reaching luminosities of $10^{41} \ \rm erg/s $ in the canonical X-ray band alone (0.3 - 10 keV; \citealt{Kaaret_2017,Pinto_2023} for a recent review). The high luminosities exhibited by these sources were originally thought to arise from accretion onto black holes with masses of up to $10^{4-5} \rm \ M_\odot$. This scenario was challenged by the discovery of pulsations in a few ULXs (pulsating ULXs, pULXs), revealing that at least $\sim 20 \% $ of ULXs are thought to be powered by accreting neutron stars (NSs; \citealt{Bachetti_2014,Israel_2016,Rodriguez_2020}). In order to explain super-Eddington luminosities in NSs, a correlation between a very high strength of the NS magnetic field (at least $\sim 10^{13} \rm \ G$), the luminosities and the mass accretion rate has been invoked (\citealt{Mushtukov_2015}). The strong magnetic field would diminish the electron scattering cross-section favouring super-Eddington luminosities. M82 X-2 and M51 ULX-7 are two sources for which such high magnetic fields values ($\sim  10^{13-14} \ \rm G$) have been suggested (\citealt{Bachetti_2020,Vasilopoulos_2020}). However, several properties characterising the pULXs can be explained by a lower magnetic field strength, ranging between $10^{11-13} \rm \ G$ (\citealt{King_2019}). The
main example is NGC 300 ULX-1, where \citet{Walton_2018b} tentatively detected a cyclotron line in this source, which suggests a magnetic field of $10^{12} \ \rm G$ (see also \citealt{Vasilopoulos_2019}). The presence of a relatively low magnetic field is also supported by the detection of a 0.2c ultrafast outflow (UFO; \citealt{Kosec_2018b}).
Moreover, the radiative mechanism of the NS accretion column at super-Eddington accretion rates and its geometry are still unknown (\citealt{West_2017}). The distance of the very few persistent pULXs and the limited capabilities of the current X-ray observatories prevent a detailed study of these sources. 

Among Galactic high-mass X-ray binaries (XRBs), we may identify two distinct types of system: persistent and transient, such as an X-ray transient accreting from an Oe/Be-type main sequence donor in an eccentric binary. One source of this last XRB class is RX J0209.6-7427, which offers the chance to improve our knowledge of pULXs for two main reasons: a) it is a nearby source that recently, in 2019, underwent an outburst, and, by following its evolution, we can gain insight into the radiative processes in the accretion column and the formation of the shock above the NS surface, which is believed to define the transition between subcritical and supercritical accretion regimes; b) this is a highly variable source located in the outer wing of the Small Magellanic Cloud (SMC) at a distance of 55 kpc. 
This system, in fact, is a Be X-ray binary system, where an accreting pulsar, with a pulse period of 9.3s (\citealt{Iwakiri_2019}), accretes matter from a donor Be star that loses material through a slow moving equatorial wind (or decretion disc, see \citealt{Reig_2011}). If the binary plane is misaligned with this equatorial wind or the NS orbit is highly eccentric, the mass transfer may be highly variable, resulting in these powerful outbursts during which the luminosity of the source surpasses the Eddington limit. During these episodes, RX J0209.6-7427 may enter an accretion regime analogous to that observed in ULXs. Powerful winds are expected in super-Eddington accretion flows, and mildly relativistic outflows have been spectroscopically detected in several ULXs (\citealt{Pinto_2016,Pinto_2017,Kosec_2018a,Pinto_2020b}). However, among confirmed pULXs, direct evidence for a relativistic outflow has so far been reported for NGC~300 ULX-1 (\citealt{Kosec_2018b}) and in the transient Be/X-ray binary A0538-66 \citep{Ducci_26}.

In this paper, we present the high-resolution X-ray spectroscopy of RX J0209.6-7427 in its super-Eddington regime. We search for spectroscopic signatures of an outflow analogous to those detected in the broader ULX population, taking advantage of its spectrum with high statistics. The paper is structured as follows: in Sect. \ref{sec:data_reduction} we report on
the observation of the source and the spectral modelling in Sect. \ref{sec:spectral_analysis}. In Sect. \ref{sec:discussion} we discuss the results of the analysis and the conclusion in Sect. \ref{sec:conclusion}.

\begin{center}
	\begin{table}[t]
\caption{XMM-\textit{Newton} observation of RX J0209.6-7427.}  
 \renewcommand{\arraystretch}{2.}
 \small\addtolength{\tabcolsep}{-3pt}
	\centering
	\scalebox{0.73}{%
	\begin{tabular}{lcccccc}
    \toprule
    {{ Obs. ID}}  &
    {{Date}}  &
    {{T$\,_{\rm net, \, RGS1}$ [s]}} &
     {{CR$\,_{\rm net, \, RGS1}$ [c/s]}} &
    {{{{T}$\,_{\rm net, \, RGS2}$ [s]}}} &
         {{CR$\,_{\rm net, \, RGS2}$ [c/s]}} \\
    \hline
0854590501 & 2019-12-22  & 11387 & 4.25 $\pm$ 0.02   & 11278 & 4.88 $\pm$ 0.02   \\
\hline
    \end{tabular}}\label{table:observations log}
       \vspace{0.3cm}
      \tablefoot
      {T$_{\rm net}$ is the exposure time after the removal of high solar flares and {CR}$\,_{\rm net, \ RGS 1/2}$ are the net source count rates.}

\end{table}
\end{center}
\section{Data reduction}
\label{sec:data_reduction}
 XMM-\textit{Newton} observed RX J0209.6-7427 (hereafter RX J0209.6) only once in 2019, when the source was in outburst, although the observation is very short (18.6 ks, PI: Schartel). Table \ref{table:observations log} contains the details of the XMM-\textit{Newton} observation that we analysed. The data analysis was performed using the XMM-\textit{Newton} Science Analysis System (SAS\footnote{\url{https://www.cosmos.esa.int/web/xmm-newton/sas}}) version 21.0.0 and the calibration files of April 2024. We reduced the data from the EPIC cameras (in timing mode and negligible pile-up) and RGS gratings following the standard pipeline in SAS (\texttt{epproc} and \texttt{rgsproc}, respectively). 
 
\begin{figure}
       \centering
       \includegraphics[width=1.\linewidth]{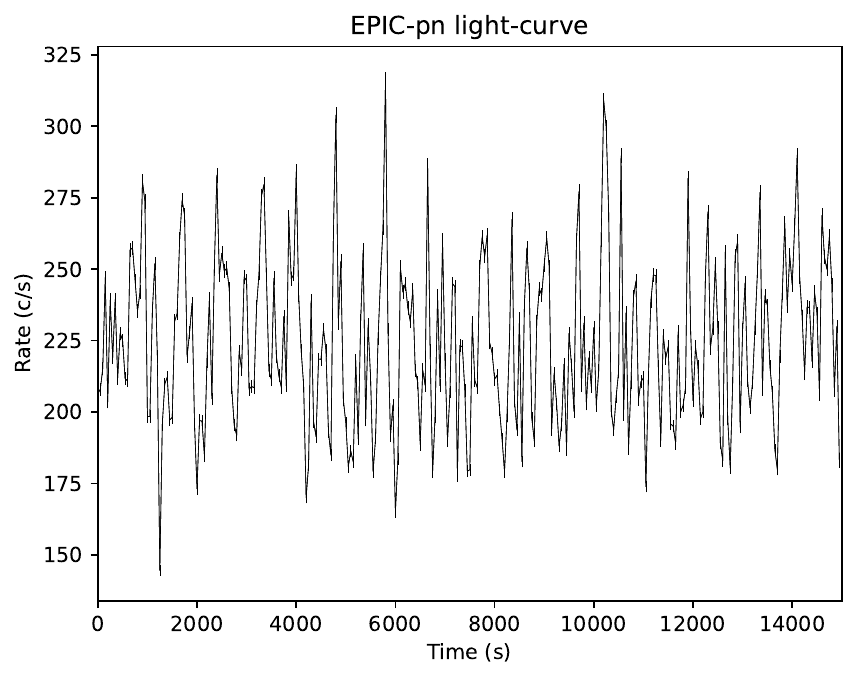}
       \caption{XMM-\textit{Newton} EPIC-pn light-curve of RX J0209.6 in the 0.3-10 keV energy band, with time bins of 50 seconds. }
       \label{fig:XMM lightcurve}
       \end{figure}

Only single- and double-pixel events (PATTERN $\leq 4$) were retained for the EPIC-pn analysis. The source extraction region was defined as a rectangular box centred on RX J0209.6 (RAWX coordinates between 33 and 43). The background was extracted from a set of columns on the same CCD, chosen to lie as far from the source region as possible. The 0.3–10 keV EPIC-pn light curve (Fig.~\ref{fig:XMM lightcurve}) shows a clear variability of the source on very short timescales ($\sim$ 100 s), while the overall source flux remains approximately constant over the full observation.

For the RGS data, we extracted first-order RGS 1/2 spectra from events within a 0.8'-wide cross-dispersion region centred on the source coordinates (RA = 02h 09m 33.85s, DEC = -74° 27' 12.5"). Background spectra were obtained by selecting photons from regions lying outside 98$\%$ of the source point-spread function, chosen to avoid overlap with bright neighbouring sources. Time intervals affected by enhanced background due to solar flares were excluded by inspecting the RGS 1/2 CCD9 (E $\ge$ 1.7 keV) light-curves and retaining only intervals with a count rate below 0.2 counts/s.

\section{Spectral analysis}
\label{sec:spectral_analysis}
The {\sc{spex}} fitting package (v3.07.03, \citealt{kaastra2023_spex}) includes advanced optically thin plasma models, which can offer additional insight into the nature of the spectral lines. By default in {\sc{spex}}, we state all uncertainties at 68\% confidence level and applied the recommended optimal binning to the spectra \citep{Kaastra_2016} for the EPIC-pn + RGS data. The Cash statistic was used for the entire analysis (\cstat, \citealt{Cash_1979}). All spectral models accounted for absorption by both the circumstellar and interstellar medium using the \texttt{hot} model, with a gas temperature fixed at $10^{-6}$ keV to describe neutral gas in {\sc{spex}}. We further adopted the most up-to-date solar abundances \citep[][the default in {\sc{spex}}]{Lodders2009}.

\subsection{Continuum spectral modelling}
\begin{figure}
    \centering
    \includegraphics[width=1.\linewidth]{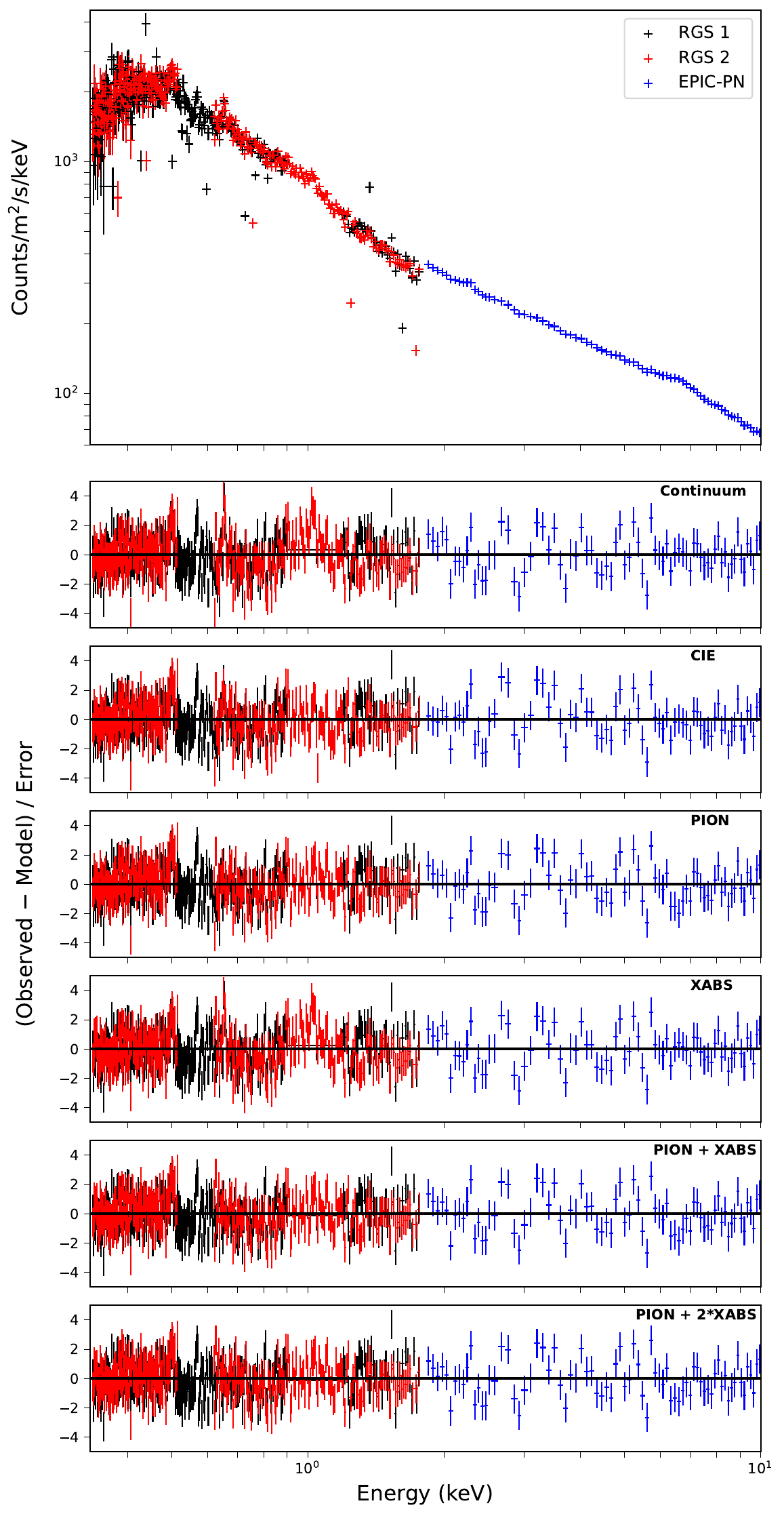}
    \caption{RGS 1/2 (in black and red) and EPIC-pn (in blue) data along with the residuals depending on the model applied. The RGS 1/2 data are regrouped by a factor of 8 directly in {\textsc{spex}} for plotting purpose. Plot residuals, from top to bottom: continuum; continuum + \texttt{cie}; continuum + \texttt{pion}; continuum + \texttt{xabs}; continuum + \texttt{pion + xabs}; continuum + \texttt{pion + 2*xabs}.}
    \label{fig:best-fit}
\end{figure}

In order to describe the RX J0209.6 X-ray spectrum, we fit the XMM-\textit{Newton} observation with a phenomenological ULX model composed of two modified blackbodies (\texttt{mbb}) and a Comptonisation (\texttt{comt}) component (see \citealt{Kosec_2021}). These take into account the emission coming from the inner ($\sim$ 1 keV) and outer ($\sim$ 0.26 keV) part of the disc (\texttt{mbb}) and the emission of the accretion column (\texttt{comt}). Furthermore, two Gaussian lines (\texttt{gaus}) were added to the continuum model, taking into account the residuals at 2.26 $\pm$ 0.01 keV (FWHM: $ 0.05 \rm \ keV$, fixed) and 6.63 $\pm$ 0.02 keV (FWHM: $\sim$ 0.91$_{-0.02}^{+0.10}$ keV), due to instrumental calibration and likely reflection effects of hard X-ray photons by the accretion disc, respectively.  The $\rm N_{H}$ of the interstellar medium was left free to vary, providing a best-fit value of (0.58 $\pm$ 0.01)$\times 10^{21} \ \rm cm^{-2} $, in agreement with that found by \cite{Kosec_2021}. For the \texttt{comt} component, coupling the seed photons with the temperature of the hotter blackbody, the fit provides an optical depth of $\sim 10$ and an electron temperature of 3.6 keV (comparable to those obtained from \citealt{Kosec_2021}, except for the seed-photon temperature). The luminosity of the source is greater than $ 4\times 10^{38} \ \rm erg/s$ in the 0.3-10 keV band alone (bolometric luminosity of $\sim 10^{39} \ \rm erg/s$) and so surpassing the Eddington limit forecasted for a NS. The model provides a reasonable description of the broadband spectrum (\cstat $\rm /d.o.f = 1559/1053 $) with the exception of the strong residuals, in emission, at $\sim$ 0.56 keV (O {\scriptsize VII}), 0.65 keV (O {\scriptsize VIII}), and 1 keV (Ne {\scriptsize IX-X} and Fe {\scriptsize L}).
In absorption, instead, a complex of spectral features is detected between 0.71 and 0.83 keV (14.5-17 Å), whose origin may be attributed either to a blueshifted blend of O {\scriptsize{VIII}} and Fe L transitions, or to rest-frame interstellar medium (ISM) and dust absorption. 
The results of the best-fit along with the residuals are reported in Table \ref{Table:Results spectral fits} and Fig. \ref{fig:best-fit}, respectively. 
A zoom of the RGS residuals is shown in Fig. \ref{fig:gaussian_line_scan} (top panel). The spectral features will be analysed in more detail in the following sections.
From this point onwards, we focused on the X-ray spectral analysis of the RGS data (0.33-1.77 keV) to search for relativistic outflows, as expected during the super-Eddington regimes.

\subsection{Gaussian line scan}
\begin{figure}[h!]
    \centering
    \includegraphics[width=.98\linewidth]{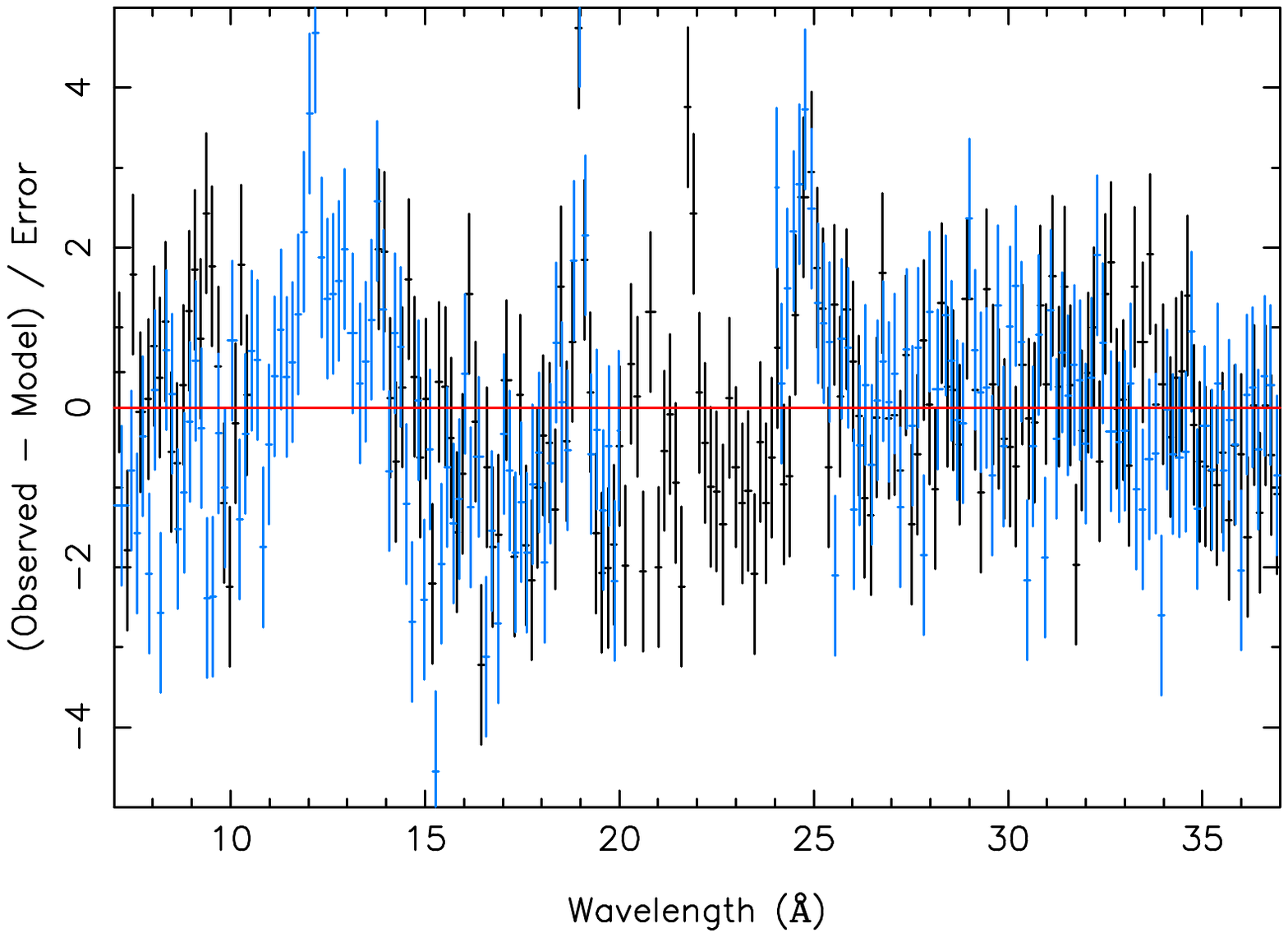}
    \includegraphics[width=1.\linewidth]{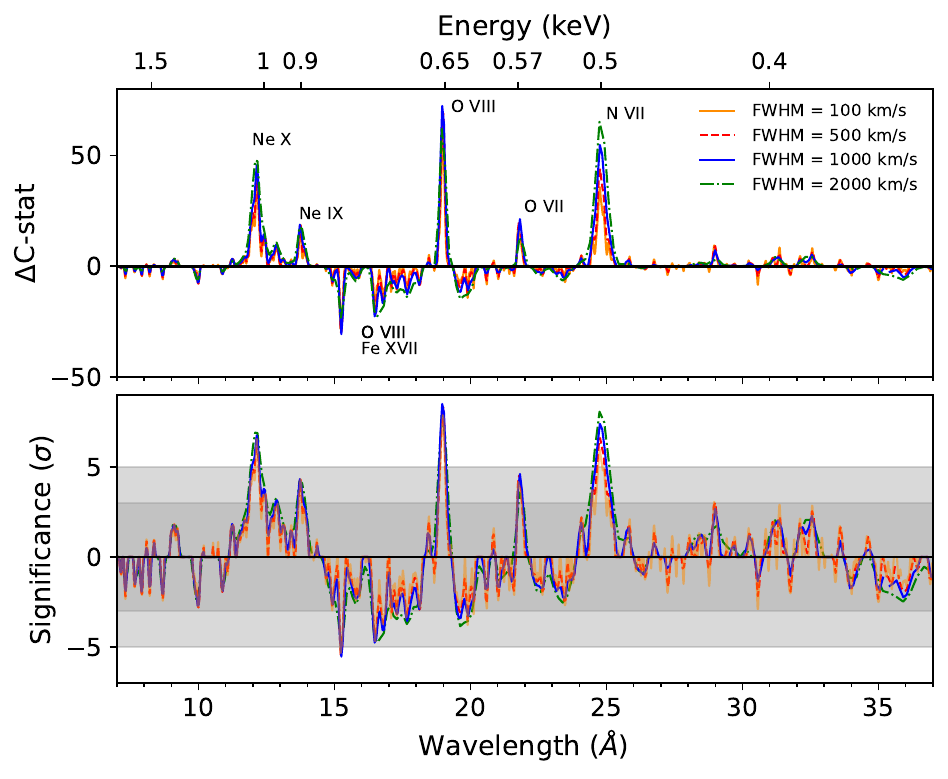}
    \caption{RGS 1/2 spectral residuals, shown in black and blue respectively, after best-fit continuum modelling. Data are regrouped by a factor of 15 in \textsc{spex} for display purposes only (top panel). Gaussian line scan with different line widths for the RGS data (7-37 \AA). The dark and light grey-shaded areas indicate the 3$\sigma$ and 5$\sigma$ single-trial significances, respectively (bottom panel).   }
    \label{fig:gaussian_line_scan}
\end{figure}

From the fitting of the data with the best-fit continuum model some residuals are present, as shown in Fig. \ref{fig:best-fit} (first bottom panel), both in emission and absorption, particularly below 2 keV. To identify and more clearly detect these spectral lines, we performed a scan of the RGS spectra with a moving Gaussian line following the same procedure applied in \citet{Pinto_2016}. We adopted a logarithmic grid with the number of points increasing from 500 to 10000 points for velocity dispersion decreasing from 2000 km/s to 100 km/s, respectively, over the 0.33 - 1.77 keV (7 - 37 Å) energy range. At each energy step we recorded the \dcstat \ improvement relative to the best-fit continuum model and derived the single-trial significance as the square root of \dcstat. This quantity was then multiplied by the sign of the Gaussian normalisation to distinguish between emission and absorption lines. The results of the line scan are shown in Fig. \ref{fig:gaussian_line_scan} (bottom panel) and Table \ref{tab:gaussian_spectral_features}.

\begin{table}[ht!]
\centering
\caption{Best-fit Gaussian line parameters for
the most significant ($\geq 3\sigma$, single-trial) spectral features.}
\label{tab:gaussian_spectral_features}

\renewcommand{\arraystretch}{1.2}
\small
\addtolength{\tabcolsep}{-1pt}

\scalebox{0.85}{%
\begin{tabular}{cccc}
\toprule
Energy centroid & FWHM & Normalisation & Identification \\
\midrule

\multicolumn{4}{c}{\texttt{Emission lines}} \\
\midrule

$0.4999 \pm 0.0008$
& $(1.04 \pm 0.21)\times 10^{-2}$
& $3.08 \pm 0.42$
& N {\scriptsize VII}$\alpha$ \\

$0.5686 \pm 0.0007$
& $(3.61 \pm 1.25)\times 10^{-3}$
& $1.36 \pm 0.33$
& O {\scriptsize VII} (triplet) \\

$0.6533 \pm 0.0005$
& $(5.66 \pm 1.32)\times 10^{-3}$
& $1.68 \pm 0.23$
& O {\scriptsize VIII}$\alpha$ \\

$0.9026  ^{+0.001}_{-0.044}$
& $(5.11^{+21.7}_{-4.85})\times 10^{-3}$
& $0.61_{-0.28}^{+0.52}$
& Ne {\scriptsize IX} (triplet) \\

$1.01 \pm 0.01$
& $0.13 \pm 0.03$
& $4.93 \pm 0.93$
& Ne {\scriptsize X} \\

\midrule
\multicolumn{4}{c}{\texttt{Absorption lines}} \\
\midrule

$0.738 \pm 0.001$
& $(5.93 \pm 3.45)\times 10^{-3}$
& $-0.49 \pm 0.16$
& O {\scriptsize VII}$\beta$ ($0.1c$) \\

$0.7516 ^{+0.0010}_{-0.0005}$
& $(1.56^{+61.8}_{-1.56})\times 10^{-4}$
& $-0.46 \pm 0.11$
& O {\scriptsize VIII}$\alpha$ ($0.1c$) \\

$0.8131 \pm 0.0009$
& $(4.38 \pm 2.03)\times 10^{-3}$
& $-0.65 \pm 0.14$
& O {\scriptsize VIII}$\alpha$ ($0.2c$) \\

\bottomrule
\end{tabular}%
}

\tablefoot{
The energy centroids and full widths at half maximum (FWHM) are given in keV, while the Gaussian normalisations are expressed in units of $10^{44}$~ph~s$^{-1}$. The absorption-line identifications are tentative.} 

\end{table}
The strongest emission features detected, with a single-trial significance $\gtrsim$ 5$\sigma$, are compatible with Ne {\scriptsize X} (12.1 Å), Ne {\scriptsize IX} ($\sim$13.5 Å), O {\scriptsize VIII} (19 Å), O {\scriptsize VII} triplet (22 Å), and N {\scriptsize VII} (24.8 Å). In absorption, we detected, with a significance $\gtrsim 4 \sigma$, a broad complex of lines between 14.5 and 17 Å, with the most prominent line at 15.23 Å ($> 5\sigma$). These lines can be interpreted either as a blend of O {\scriptsize VIII} and Fe {\scriptsize XVII–XVIII} with velocities of $\sim$0.1c (see Sect. \ref{sec:discussion}), as also suggested by \citealt{Kosec_2021}, or as arising in the ISM. However, the individual absorption-line identifications remain tentative. In particular, the interpretation of the feature near 0.813 keV as O {\scriptsize VIII}$\alpha$ from an absorber moving at approximately $-0.2c$ is non-unique, since it is driven predominantly by a single absorption feature. Alternative ionic identifications or blends cannot be excluded.

For the candidate absorber at approximately $-0.1c$, identifying the feature at $0.738 \pm 0.001$ keV with O {\scriptsize VII}$\beta$ implies a corresponding He$\alpha$ transition at approximately 0.636 keV (19.48~\AA). The Gaussian scan shows an absorption-like residual near this position, which may represent a weak counterpart. However, consistency with the velocity and relative line strength expected from the same absorber has not been established. The absence of a securely identified He$\alpha$ counterpart therefore remains an important caveat to the proposed He$\beta$ identification. The near-rest-frame O {\scriptsize VII} emission triplet is spectrally separated from the predicted blueshifted He$\alpha$ position and cannot fill this absorption feature.
To investigate a possible interstellar origin, we initially modelled these absorption features by adding the \texttt{amol} model, which accounts for absorption by interstellar dust.
The \texttt{amol} model calculates the transmission of various molecules considering both absorption and scattering. We left free to vary the molecular column density and fixed the molecular number parameter to 126 (metallic iron in {\textsc{spex}}). The improvement of the fit, by adding this component, is relatively poor (\dcstat $\sim$ 10) with respect to the continuum model. Furthermore, we fitted the spectrum by adding a \texttt{hot} component to the continuum model, with an input temperature of 0.15 keV, in order to describe absorption features of Galactic highly-ionised ISM plasma, as performed by \citet{Pinto_2013}. The fit, also in this case, provided a not very significant improvement with respect to the continuum model (\dcstat $\ \sim$ 10), slightly improving the fit of the O {\scriptsize VII} resonance line at 21.6 Å. Neither of the tested ISM configurations fully reproduces the broad absorption complex. These tests therefore disfavour the specific interstellar absorption models considered here, but do not establish a unique physical origin for the features.
In order to understand the origin of the latter lines, we performed a multidimensional scan using optically-thin plasma models.

\subsection{Multi-dimensional grids with photo-/collisionally-ionised plasma}
Plasma detection becomes more reliable when multiple spectral lines are considered at once. For this reason, we carried out a broad search of the multi-dimensional parameter space for plasma in photo-ionisation equilibrium (PIE) and collisional ionisation equilibrium (CIE). This method helps to avoid one of the main problems of line fitting—the tendency for solutions to get stuck in local minima when fitting single lines or starting from arbitrary initial guesses. It also makes it possible to identify several components or phases within the plasma, which is key to determining its true properties. Although this approach is computationally intensive, we improved efficiency by adjusting model parameters step by step and making use of pre-calculated model grids. The contour plot lines shown in the grids indicate the significance of the solutions at 3,4,5 $\sigma$ as estimated from Monte-Carlo simulations, taking into account for the look-elsewhere effect (see \citealt{Pinto_2020b}).

\subsubsection{Collisionally-ionised plasma}
\begin{figure}
    \centering
    \includegraphics[width=0.92\linewidth]{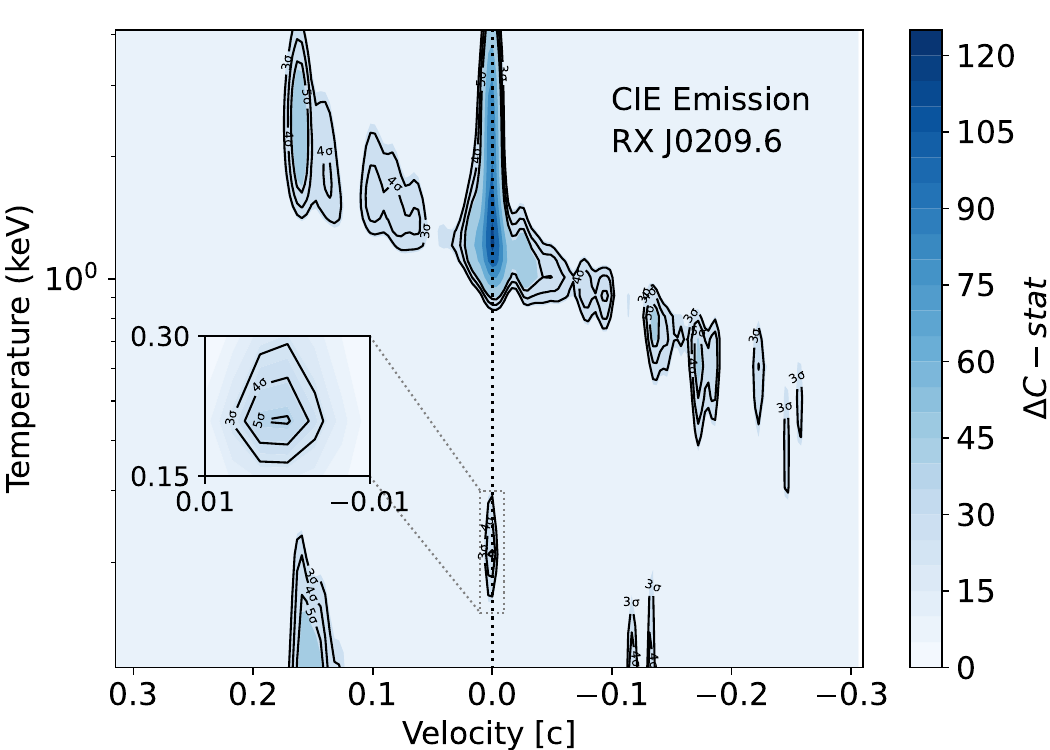}
    \caption {Grid of collisionally-ionised emitting plasma in equilibrium. The contour plot lines indicate the significance at 3,4,5 $\sigma$. The inset plot shows the secondary solution which produces the O {\scriptsize VII} lines.}
    \label{fig:cie_emission}
\end{figure}

We carried out a multi-dimensional scan using an emission model of a plasma in collisional ionisation equilibrium (\texttt{cie} model in {\sc{spex}}), following the same procedure as performed in \cite{Kosec_2018b} and \cite{Pinto_2020b}. A logarithmic temperature grid was adopted, ranging from 0.1 to 4 keV in 50 bins, together with line-of-sight velocities, $v_{\rm LOS}$, spanning from –0.3c to +0.3c (corresponding to blueshifted and redshifted plasma, respectively), in steps of 1000 km/s. The $v_{\rm LOS}$ parameter was provided by the \texttt{reds} model with {\sc{FLAG=1}}, i.e. the Doppler shift rather than the cosmological redshift, which is applied to the \texttt{cie} component.
The plasma turbulence, $v_{\rm RMS}$, was fixed at 1000 km/s. Solar abundances were assumed for the \texttt{cie} model to avoid unnecessary degeneracies and to significantly reduce computational time. The emission measure, defined as EM = $n_{e}n_{\rm H}V$, was the only free parameter in the \texttt{cie} model during the spectral fits.
The best-fit results indicated a large improvement over the continuum-only model ($\Delta$\cstat $\sim$ 125). The fit suggests the presence of an emitting, rest-frame plasma with a temperature of approximately 1.2 keV (see Fig. \ref{fig:cie_emission}).

\subsubsection{Emitting plasma in photo-ionisation equilibrium}
\begin{figure}
    \centering
\makebox[\linewidth][c]{%
  \hspace*{-12.7bp}%
  \includegraphics[width=.7762\linewidth]{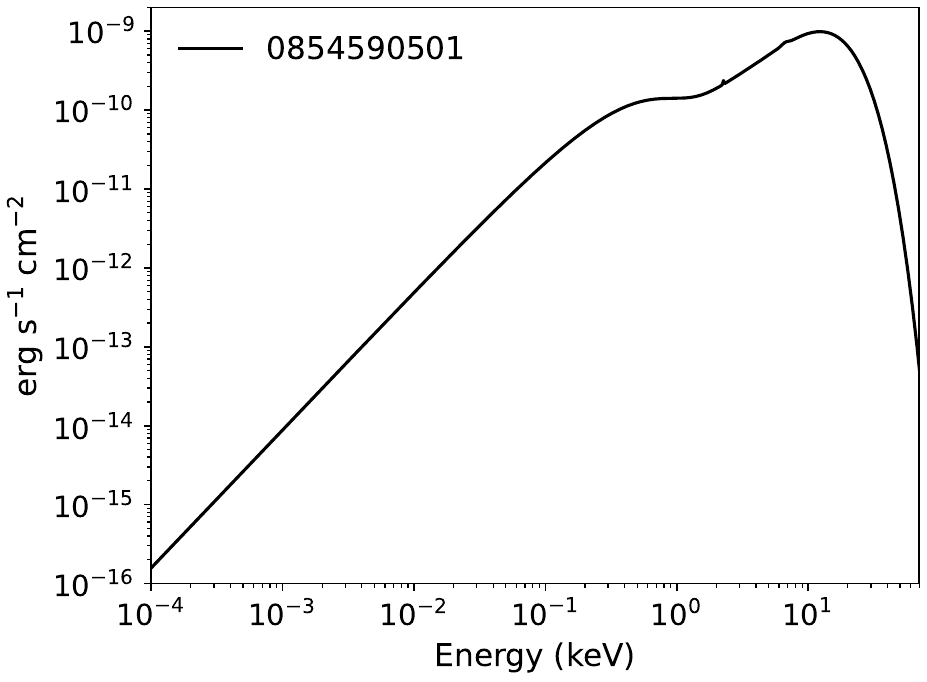}%
  \hspace*{12.7bp}%
}\par
    \includegraphics[width=.83\linewidth]{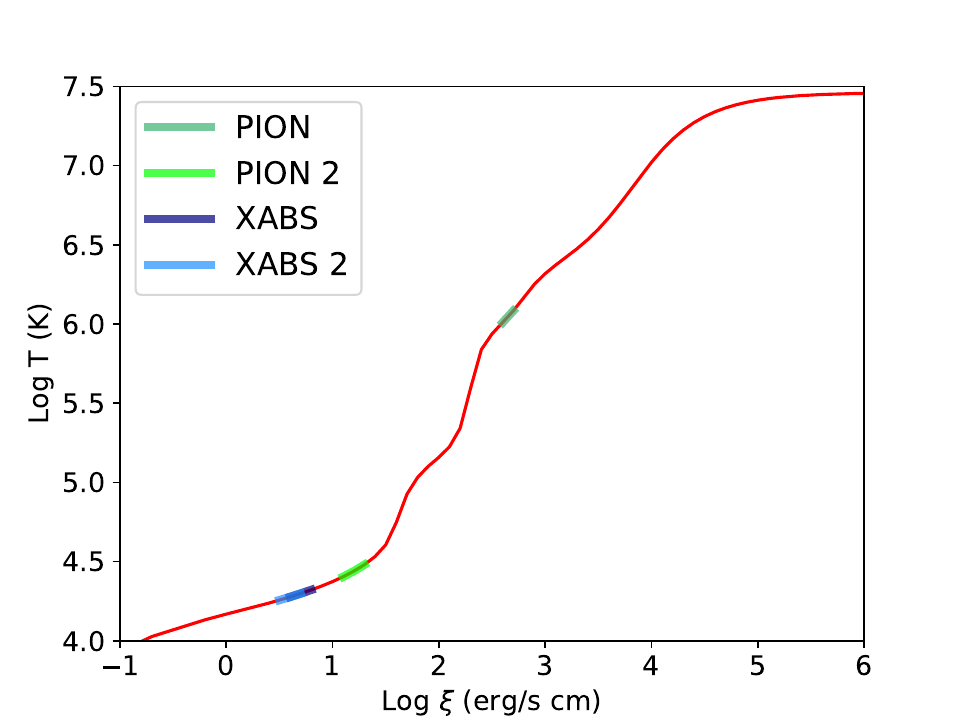}
    \includegraphics[width=.83\linewidth]{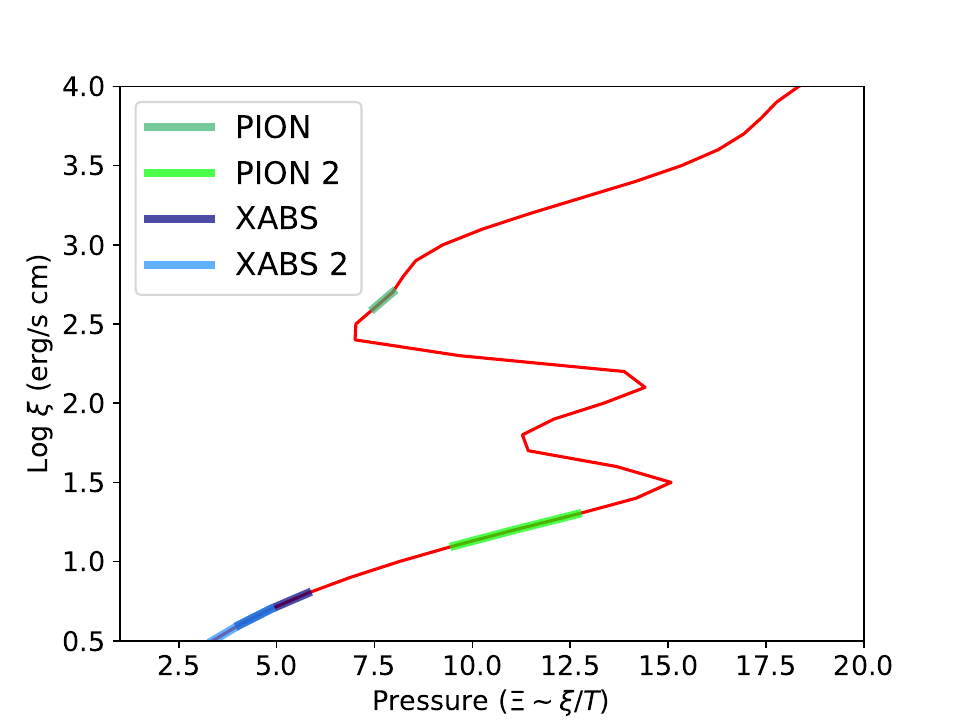}
    \caption{Spectral energy distribution from optical to the hard X-rays energy band (0.0001-100 keV, top panel), ionisation balance (middle panel) and thermal-stability curves (bottom panel) computed for the XMM-\textit{Newton} spectrum. The regions in which there are thermal instabilities are identified by the segments with negative slopes. Thicker segments show the ranges of the best-fitting solutions.}
    \label{fig:sed_stabilty_balance}
\end{figure}

These features are modelled using detailed photo-ionisation calculations as well as winds proposed to explain the 1 keV features observed in certain Galactic XRB presenting bursts (such as 4U 1820-303, \citealt{Barra_2025}) and ULXs (i.e. Holmberg II X-1, \citealt{Barra_2024}). Such models require an accurate understanding of the radiation field, specifically the spectral energy distribution (SED) spanning from optical to hard X-ray wavelengths.
The SED strongly peaks in the X-ray band, which is covered by XMM-\textit{Newton} and restrains any systematic uncertainties in the ionisation balance computation because we know only little about the other energy bands. A very important parameter of the photo-ionisation equilibrium is the ionisation parameter $\xi$ = $L_{\rm ion}$/($n_{\rm H} \ R^2$), where $L_{\rm ion}$ is the ionising luminosity (estimated\ in the 13.6 eV - 13.6 keV range), $n_{\rm H}$ is the hydrogen volume density, and $R$ is the distance from the ionising source  (\citealt{Tarter_1969}). The ionisation balance was calculated with the {\sc{spex}} \texttt{pion} model, which calculates the transmission and/or the emission of a thin slab of photo-ionised gas self-consistently. The stability (or S) curve was also computed, which is the relation between the temperature (or the ionisation parameter) and the ratio of the radiation and the thermal pressure, which can be expressed as $\Xi$ = F/($n_{\rm H}$ckT) = 19222 $\xi/T$ (\citealt{Krolik_1981}). The SED and the stability curve are reported in Fig. \ref{fig:sed_stabilty_balance}.
\begin{figure}[t!]
    \centering
\makebox[\linewidth][c]{%
    \hspace*{2.3mm}%
    \includegraphics[width=0.9\linewidth]{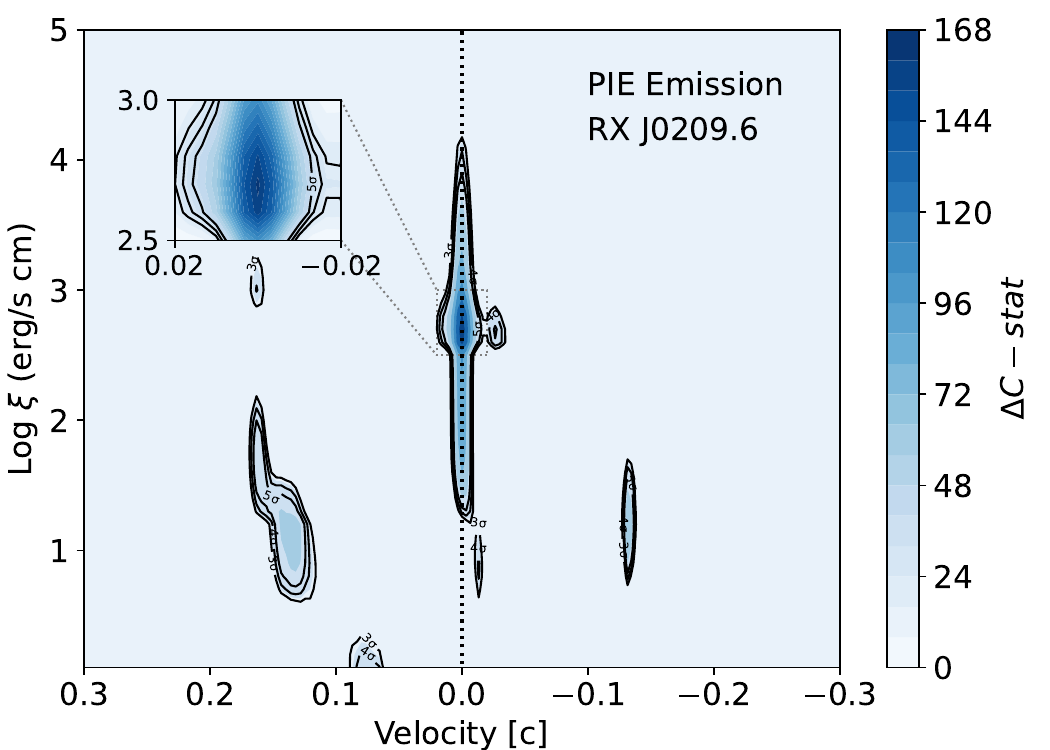}
    \hspace*{-2.3mm}%
}
    \caption{Grid scan of photo-ionised emitting plasma in equilibrium. The contour plot lines indicate the significance at 3,4,5 $\sigma$.}
    \label{fig:pion_emission}
\end{figure}

We first describe the emission lines by adding the \texttt{pion} component to the continuum model. In a similar way as for the CIE grids, we created a logarithmic grid with an ionisation parameter log $\xi$ [erg /s cm] between 0 and 5 with 0.1 steps and the $v_{\rm LOS}$ from -0.3c to +0.3c with a step of 1000 km/s, and only the \texttt{pion} column density $n_{\rm H}$ was left free to vary. The plasma turbulence $v_{\rm RMS}$, in a similar way to the CIE grid, was fixed at 1000 km/s. Therefore, we set the \texttt{pion} covering fraction $f_{\rm cov}$ to zero, such that this model exclusively produced the emission lines, and $\Omega$, the solid angle, was fixed to 4$\pi$ (1 in {\sc{spex}} units). This allowed us to save substantial computational time. This model produces a better improvement than that obtained with the \texttt{cie} component (\dcstat $\sim$155), corresponding to a rest-frame photoionised plasma solution with log $\xi \sim$ 2.6, as shown in Fig. \ref{fig:pion_emission}.

\subsubsection{Physical properties of the line-emitting plasma}
\label{sec: density_emission lines}
We investigated the physical conditions of the emitting plasma through the density- and temperature-sensitive line ratios of the He-like triplets. In
particular, following \citet{Pourquet_2000}, we considered $R = \frac{f}{i} $ and $   G = \frac{f+i}{r},$ where $r$, $i$, and $f$ are the fluxes of the resonance,
intercombination, and forbidden transitions, respectively. The $R$ ratio is primarily sensitive to the electron density, whereas the $G$ ratio can be
used to probe the electron temperature and the dominant ionisation mechanism.

\begin{figure*}[t!]
    \centering
    \includegraphics[width=.48\linewidth]{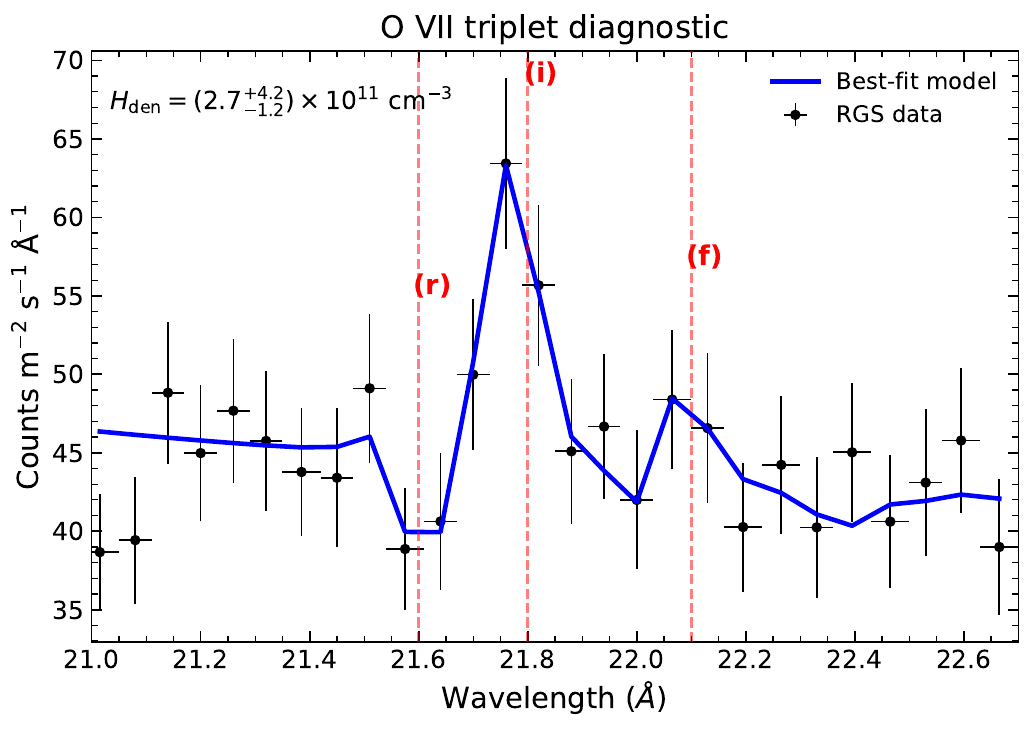}
     \raisebox{4bp}{%
  \includegraphics[width=.46\linewidth]{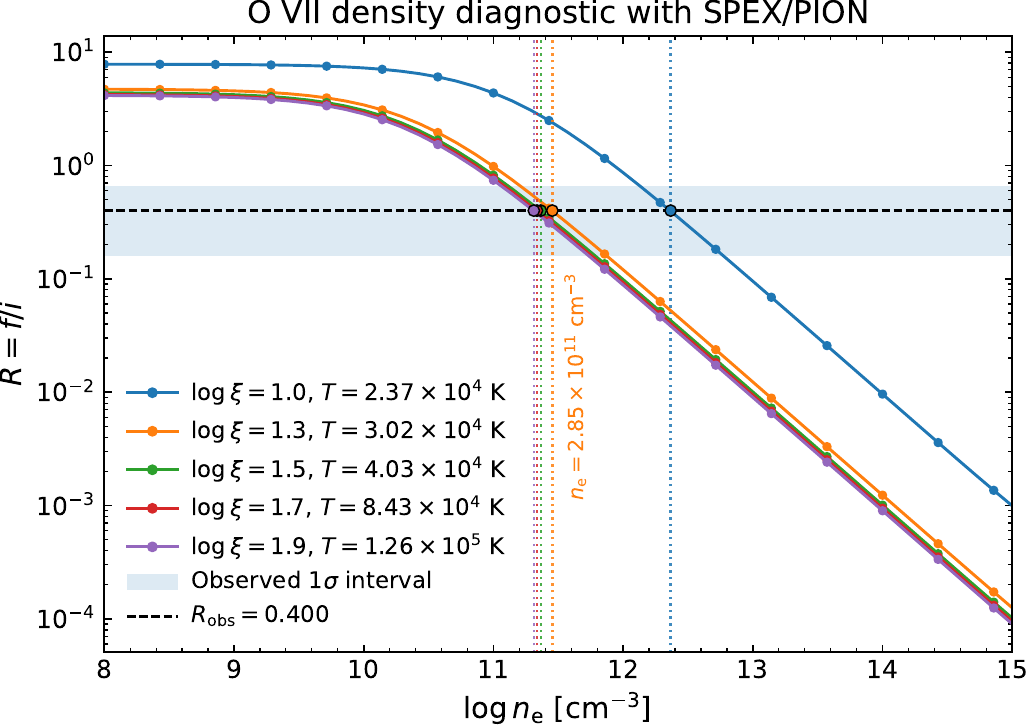}
}
    \caption{Left: Zoom on the O {\scriptsize VII} emission triplet (r: resonance, i: intercombination, f: forbidden) as fitted with three Gaussian lines in the RGS spectrum. The red-dashed lines indicate the laboratory wavelength of the O {\scriptsize VII} triplet lines. Right: R ratio curve for the O {\scriptsize VII} triplet vs the electron density. The black-dotted line indicates the observed value through the ratio  $R = \frac{f}{i} $. The grey shadow region represents the uncertainties associated with the R value.}
  \label{fig: density}
  \end{figure*}

We applied this diagnostic to the O {\scriptsize VII} triplet identified in the Gaussian line scan of the RGS spectrum (see Fig.~\ref{fig: density}, left panel). The three transitions were described using Gaussian components whose widths and velocity shifts were coupled. The fit yielded line normalisations of $r = 0.47_{-0.44}^{+0.48}$, $i = 1.95_{-0.43}^{+0.46}$, $f = 0.78_{-0.42}^{+0.47}$, in units of $10^{44}~{\rm photons\,s^{-1}}$. The best-fitting common FWHM was ($1.80 _{-1.80}^{+1.11})\times10^{-3}$ keV, while the centroids of the three lines were shifted by  $(-1.3 \pm 0.6)\times10^{-3}\rm c$ ($\sim$ 400 km/s) with respect to their laboratory wavelengths.

Using the best-fitting Gaussian normalisations, we measured $R = 0.40_{-0.25}^{+0.42}$ and $G = 5.84_{-5.84}^{+5.67}$. The $R$ ratio derived from the Gaussian components corresponds to an electron density of $n_{\rm e} = \left(2.7_{-1.2}^{+4.2}\right)\times10^{11}\ {\rm cm^{-3}}$. The $G$ ratio is only weakly constrained because the resonance-line normalisation is comparable to its statistical uncertainty. Consequently, this ratio does not allow us to distinguish between photoionised and collisionally ionised plasma.
To translate the observed line ratios into constraints on the plasma properties, we calculated the expected O {\scriptsize VII} $R$ and $G$ ratios using the \textsc{spex} \texttt{pion} model. We  adopted the best-fitting continuum model as the incident spectral energy distribution and constructed a grid over the relevant plasma parameters. At each grid point, the ionisation balance and the luminosities of the O {\scriptsize VII} triplet lines were extracted using the \texttt{ascdump} command.
The calculations covered the ionisation-parameter range $1.0 \leq \log\xi \leq 1.9$, with a step of $0.2$, and electron densities between $10^{8}$ and $10^{15}\ {\rm cm^{-3}}$, sampled with a logarithmic spacing of $\Delta\log n_{\rm e}=0.1$. We assumed a velocity dispersion of $1000\ {\rm km \ s^{-1}}$ and a hydrogen-equivalent column density of the order of $10^{20}\ {\rm cm^{-2}}$, consistent with the values producing the largest improvement in the spectral fit obtained in Table \ref{Table:Results spectral fits}.
The measured value $R=0.40$ intersects the theoretical curves at an electron density of approximately $n_{\rm e}\sim 3 \times10^{11}\ {\rm cm^{-3}}$. By considering the uncertainties associated with the R value, the inferred density spans approximately $\rm 10^{11-12} \ cm^{-3}$. The results are reported in Fig. \ref{fig: density} (right panel). 
We repeated the same analysis for the N {\scriptsize VI} triplet. In this case, the measured line ratio yields a lower limit on the electron density of $n_{\rm e}>2\times10^{11}\ {\rm cm^{-3}}$. The constraints obtained from the O {\scriptsize VII} and N {\scriptsize VI} triplets are therefore mutually consistent and indicate that the emission originates from a relatively dense plasma. The inferred electron density is also comparable to that reported for the ULX NGC 1313~X-1 \citep{Pinto_2020a} and NGC 5204 X-1 \citep{Caserta_2026}.
The secondary emission solution associated with the O {\scriptsize VII} triplet can be described either by a \texttt{cie} component with a temperature of $kT\sim0.2\ {\rm keV}$, as shown in the inset of Fig.~\ref{fig:cie_emission}, or by a \texttt{pion} component with an ionisation parameter of $\log\xi\sim1.3$ (see Fig. \ref{fig:1_PMF_absorber}). Motivated by the triplet diagnostics, we repeated the photoionised emission-line modelling using spectra calculated with \texttt{pion} at a fixed hydrogen number density of $n_{\mathrm H}=3\times10^{11}\,\mathrm{cm^{-3}}$ and a velocity dispersion of $1000\,\mathrm{km\,s^{-1}}$. The corresponding emission grids and spectral fits are presented in Appendix \ref{appendix-number_density}, together with the collisionally ionised plasma scans (see Fig. \ref{fig:cie_emission_new_bin_hden}). The photoionised models provide a better fit to the O {\scriptsize VII} complex in the configurations explored here, which also include a near-rest-frame absorber.  
The two-component photoionised solution is characterised by ionisation parameters of $\log\xi=1.3\pm0.3$ and $\log\xi=2.7\pm0.1$, with the lower-ionisation component accounting primarily for the O {\scriptsize VII} complex. The best-fit configurations with one and two photoionised emission components are shown in Figs.~\ref{fig:1_PMF_absorber} and~\ref{fig:2_PMF_absorber}, respectively, along with the scan grids. Both include a near-rest-frame \texttt{xabs} component to describe absorption affecting the O {\scriptsize VII} resonance line. For the two-emitter configuration shown in Fig.~\ref{fig:2_PMF_absorber}, this absorber has $N_{\rm H}=(1.5\pm0.3)\times10^{20}\ \mathrm{cm^{-2}}$, $\log\xi=0.93_{-0.13}^{+0.10}$, and $v_{\rm LOS}=-320_{-260}^{+190}\ \mathrm{km\,s^{-1}}$. This near-rest-frame component is distinct from the relativistically blueshifted absorbers investigated below.

\subsection{Absorbing plasma in photo-ionisation equilibrium}
\label{subsec:Absorbing plasma in photo-ionisation equilibrium}
Finally, we performed a multi-dimensional grid scan using an absorption model. Although the \texttt{pion} model is highly versatile, it recalculates the ionisation balance at each iteration, making it computationally intensive. For the absorbing gas, we therefore adopted the more efficient \texttt{xabs} model, which is specifically optimised for absorption. This model relies on a pre-calculated ionisation balance from \texttt{pion}, thereby significantly reducing the computational load.
The \texttt{xabs} model shares most parameters with \texttt{pion}, with the exception of the opening angle of the line emission, which is fixed at zero since no emission is included. We further assumed a covering fraction of unity to minimise degeneracy and reduce computing time. The grid of photo-ionised \texttt{xabs} models was generated in the same manner as for \texttt{pion}, but with line-of-sight velocities, $v_{\rm LOS}$, ranging from –0.3c to 0 (i.e. only outflowing absorbing gas, or a wind).
\begin{figure}
    \centering
    \includegraphics[width=.95\linewidth]{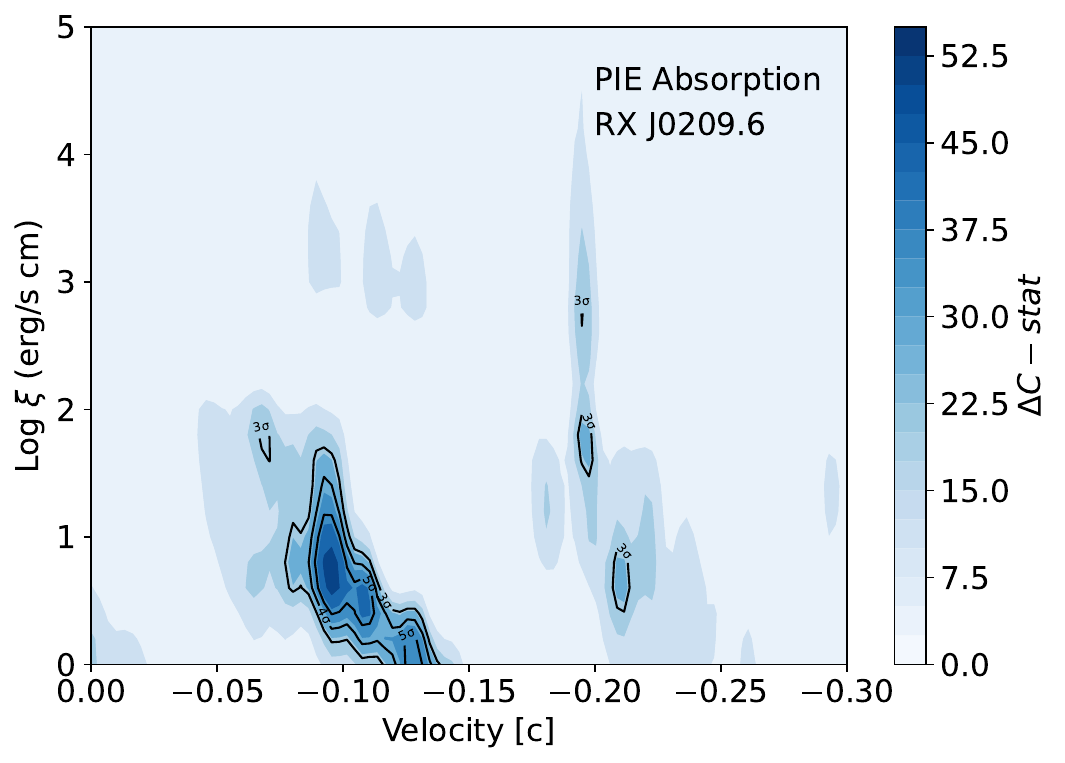}
    \centering
    \includegraphics[width=.91\linewidth]{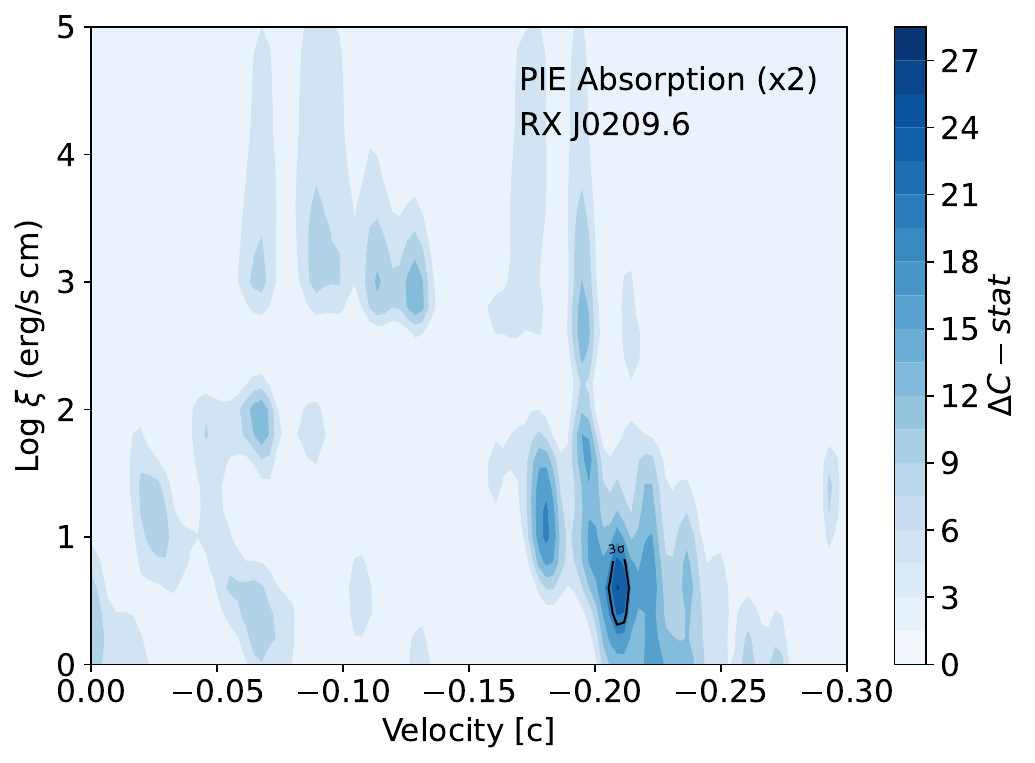}
    \caption{Grids scan of photo-ionised absorbing plasma in equilibrium with one (top panel) and two (bottom panel) \texttt{xabs} components, respectively. The bottom grids scan was produced by adding a second \texttt{xabs} component and re-running the scan starting from the best-fit solutions identified in the top grid. The contour plot lines indicate the significance at 3,4,5 $\sigma$.}
    \label{fig:pion_absorption}
\end{figure}
The best-fit solution corresponds to an improvement of \dcstat $\sim$ 58 (shown in Fig. \ref{fig:pion_absorption}) relative to the continuum-only model, with $\log \  \xi \sim 0.8$ and a velocity $v_{\rm LOS} \sim -0.1$c, indicating an absorbing, blueshifted plasma scenario. Compatible results are obtained with \texttt{pion} in absorption (see Table \ref{Table:Results spectral fits}).

The grid search also identifies a second candidate absorption solution (\dcstat $\sim$ 30 relative to the continuum + \texttt{xabs}, see Fig. \ref{fig:pion_absorption}, bottom panel), suggesting a possible second absorbing component with  $\log \  \xi \sim 0.6$ moving at $v_{\rm LOS} \sim -0.2$c, similarly to those expected from archetypal ULXs in super-Eddington regimes (e.g. NGC 1313 X-1 \citealt{Pinto_2020b}). These exploratory results depend on the adopted description of the emission spectrum.

To assess the impact of the emission-line modelling on the inferred absorption properties, we first modelled the emission features using two \texttt{pion} models, adopting the plasma density discussed in Sect. \ref{sec: density_emission lines}, and the rest-frame absorber for taking into account the O {\scriptsize VII} resonance line absorption. We then repeated the \texttt{xabs} grid searches with these emission components included in the spectral model. The resulting grids are shown in Figs. \ref{fig:xabs_emission_density_correct_1xabs} and \ref{fig:xabs_emission_density_correct_2xabs}.
The slower absorbing component is recovered at approximately $0.1c$, with a fit improvement, with respect to the model presented in Sect. \ref{sec: density_emission lines}, of \dcstat $\simeq 24$ ($\gtrsim3\sigma$). Adding a further absorbing component yields an improvement of \dcstat $\simeq 15$ ($\sim2.5\sigma$), with a preferred velocity of approximately $0.2c$. The evidence for this second, faster component therefore remains tentative.

\subsection{Collisionally ionised absorber}
\label{sec:hot_absorber}

As an alternative to the blueshifted photoionised absorbers, we tested a collisionally ionised absorber at rest. We added an additional \texttt{hot} component in \textsc{spex} to the model with two \texttt{pion} emission components, retaining the neutral absorption included in all models. The absorber temperature and column density were left free, while the line-of-sight velocity was fixed at $v_{\rm LOS}=0$ and the turbulent velocity dispersion at $v_{\rm RMS}=100$~km~s$^{-1}$. The fit yields $kT=0.019\pm0.002$~keV and $N_{\rm H}=(0.14\pm0.02)\times10^{21}$~cm$^{-2}$. Adding this component improves the fit by \dcstat=13 for two additional free parameters, giving \cstat$/\mathrm{d.o.f.}=1307/1042$ (see Table \ref{Table:Results spectral fits}). The comparison with the photoionised absorption models is discussed in Sect.~ \ref{subsec: Constraints on the emission line properties}.
\section{Results and Discussion}
\label{sec:discussion}
This paper focused on the analysis of the XMM-\textit{Newton} spectrum of the pULX RX J0209.6 during its outburst. At first, we analysed the continuum by adopting a phenomenological model used for the ULXs. Then, we conducted a scan search for spectral lines with the optically-thin plasma models provided by {\sc{spex}}.
Before searching for lines, we needed to describe the continuum spectrum. The best-fit continuum model (\cstat/d.o.f. $1559/1053$) is the following: \texttt{hot*(mbb+mbb+comt+2*gaus)}. This model, composed of three continuum components, generally used for the description of the ULXs spectra, describes the emission coming from the inner (outer) and hot (cold) parts of the accretion flow. Moreover, the emission of the accretion column was taken into account with the \texttt{comt} component.  
In the end, two Gaussian lines take into account the residuals at 2.2 and 6.63 keV, due to instrumental calibration and reflection-like feature, respectively. The results are shown in Fig. \ref{fig:best-fit} and Table \ref{Table:Results spectral fits}. However, some residuals are still present at $\sim$ 0.5 keV, 0.56 keV, 0.65 keV and 1 keV in emission and some in absorption at $\sim$ 0.7 - 0.8 keV. We, therefore, proceeded to the modelling of these lines with state-of-the-art optically-thin models provided by {\sc{spex}} and for searching for relativistic outflows as expected by the super-Eddington regime. 
\subsection{Spectral lines}
To assess the significance of any plasma component, we fitted the continuum and then performed a comprehensive multi-dimensional scan of equilibrium plasma models, using \texttt{pion/xabs} for photo-ionised emission/absorption and \texttt{cie} for collisional ionisation. The results are provided by Table \ref{Table:Results spectral fits} and the fit residuals are shown in Fig. \ref{fig:best-fit}. Using the \texttt{cie} component, the grids scan in Fig. \ref{fig:cie_emission} suggested an emitting, rest-frame plasma with a temperature of $\sim$ 1.25 keV. The \cstat/d.o.f provided by the \texttt{cie} modelling is 1434/1050 (\dcstat $\sim$ 125 with respect to the continuum only modelling). 

We also attempted to describe the emission lines with a photo-ionised emitting plasma component. The results of the grids scan, reported in Fig. \ref{fig:pion_emission}, suggested also in this case an emitting, rest-frame plasma with a log $\xi$ of 2.6 and a hydrogen column density of (3.5 $\pm$ 0.5) $\times$ $10^{21} \rm \ cm^{-2}$, improving the fit by \dcstat$\sim 155$ relative to the continuum-only model and by \dcstat$\sim 30$ relative to the continuum+\texttt{cie} model.

Although both two emission models are confirming a rest-frame solution, the larger improvement provided by the \texttt{pion} component suggests a photo-ionised plasma-like rather than a shocked one. We notice that the \texttt{pion} grids identify two further solutions at $|v|\gtrsim0.12c$ (see Fig.\,\ref{fig:pion_emission}). These solutions (which contribute with a much lower \dcstat) are due to non-realistic, alternative descriptions of some lines. For instance, the redshifted solution with log\,$\xi\gtrsim1$ tries to model the strong (and rest-frame) N {\scriptsize VII} line with a redshifted O {\scriptsize VII} triplet; the other one refers to the O {\scriptsize VII} triplet blueshifted in the attempt of modelling the O {\scriptsize VIII} line.
We then described the absorption features using the \texttt{xabs} component, with the explored parameter grid shown in Fig. \ref{fig:pion_absorption} (top panel, after continuum modelling only). The main absorption complex lies between $0.71$ and $0.83$~keV. To investigate its origin, we performed \texttt{xabs} grid searches that identified candidate absorbers at $v_{\rm LOS}\simeq-0.1c$ and $-0.2c$ (see Sect.~\ref{subsec:Absorbing plasma in photo-ionisation equilibrium}). In the fit with two \texttt{xabs} components added to the continuum (Table~\ref{Table:Results spectral fits}), the slower candidate has $N_{\rm H}=(0.09\pm0.01)\times10^{21}$~cm$^{-2}$ and $\log\xi=0.75\pm0.07$. The faster candidate has $N_{\rm H}=(0.06\pm0.01)\times10^{21}$~cm$^{-2}$ and $\log\xi=0.62\pm0.09$.

As an intermediate step, we fitted the spectrum with one photoionised emission component and two photoionised absorption components, obtaining \cstat/d.o.f.$=1353/1044$ (\dcstat$\simeq206$ relative to the continuum-only model). The best-fitting plasma parameters are consistent, within their uncertainties, with those obtained when modelling emission and absorption separately. This configuration suggests a multiphase plasma with distinct ionisation properties and line-of-sight velocities. The best-fitting parameters are listed in Table~\ref{Table:Results spectral fits}, and the residuals are shown in the bottom panel of Fig.~\ref{fig:best-fit}.

\subsection{Constraints on the emission and absorption line properties}
\label{subsec: Constraints on the emission line properties}
\begin{figure}
    \centering
    \includegraphics[width=1.\linewidth]{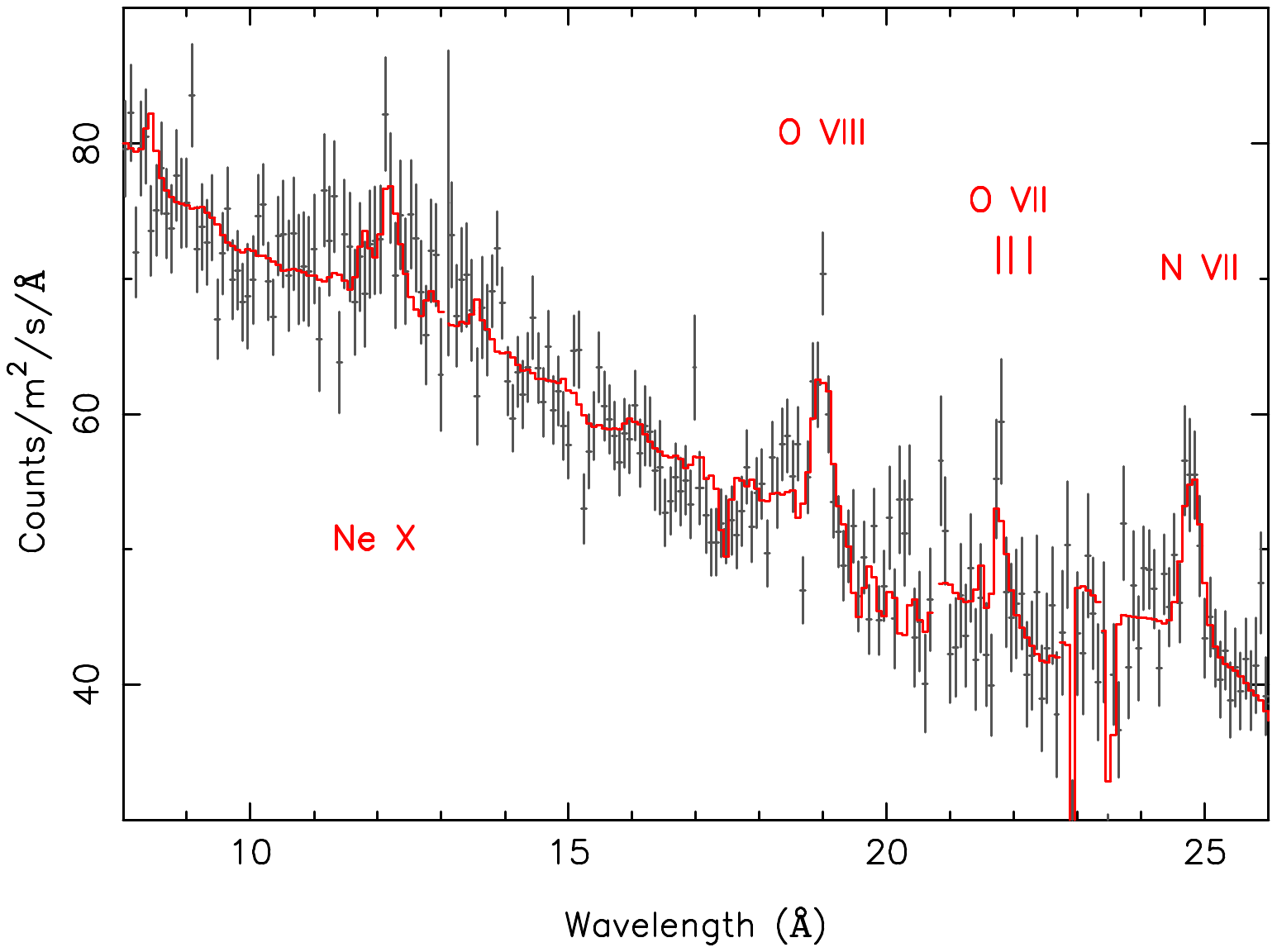}
    \caption{Spectral modelling of the RGS 1/2 data with the best-fit continuum model + 2 \texttt{cie} components. The data was rebinned for displaying purposes.}
    \label{fig:pion_emission_RGS}
\end{figure}

Strong emission features are visible in the RGS spectrum at 12.1 Å (Ne {\scriptsize X}), 19.0 Å (O {\scriptsize VIII}), 21.8 Å (O {\scriptsize VII} triplet complex), and 24.7 Å (N {\scriptsize VII}), as shown in Fig. \ref{fig:pion_emission_RGS}. 

By modelling these lines with optically-thin plasma models, it is possible to constrain several properties of the emitting plasma, such as its number density $\rm h_{den}$. 
As seen in Fig. \ref{fig:best-fit}, these emission lines cannot be fully  described with a single emission component. The application of two \texttt{pion} components in emission is extremely slow, especially from a random set of abundances. We therefore first attempt a fast modelling with two \texttt{cie} components as follows.
Two \texttt{cie} components, then, were introduced, which are computationally faster than the \texttt{pion} plasma model, and added on top of the best-fitting absorption model (continuum + \texttt{xabs}). In addition, a \texttt{reds} component was included to account for the velocity of the emitting plasma, as previously done for the \texttt{cie} grids scan. Such model can be described as follows: \texttt{hot*(xabs*(mbb+mbb+comt)+reds*(cie+cie))}. Assuming Solar abundances, except for N, O, and Fe, all parameters of the two \texttt{cie} components were coupled, apart from their normalisations and temperatures.
The spectral fit suggests a multiphase, slowly-moving ($10\pm 200$ km/s), emitting plasma with temperatures of 0.16 $\pm$ 0.01 keV and 1.07 $\pm$ 0.05 keV. This plasma is characterised by a number density of $\rm h_{den}$ $\sim$  (0.2 $_{-0.1}^{+2.4}$) $\times 10^{11} \ \rm cm^{-3}$, in agreement, within the uncertainties with the value found in Sect. \ref{sec: density_emission lines}, and turbulent velocity  $ \rm v_{RMS}=$ 1700 $\pm$ $300$ km/s. The best-fitting abundances for the nitrogen, oxygen and iron are $\rm N=5.0\pm 0.2; \rm O=0.9 \pm 0.2; Fe= 0.4 \pm 0.1$, respectively. 
Some of these properties differ from those typically observed in ULXs. For instance, the 0.3-10 keV luminosities here, $\rm (0.010 \pm 0.003) \times 10^{38}\  erg/s$ and $( \rm 0.041 \pm 0.005) \times 10^{38} \ erg/s$, for the cool and hot plasma components respectively, are significantly lower than in persistent, extragalactic, ULXs, whereas the number density is comparable (see, e.g., for the NGC 1313 X-1 and NGC 55 ULX-1 case; \citealt{Pinto_2020b,Pinto_2025}). Finally, we fitted the RGS data using a double photoionised \texttt{pion} model in emission. In order to make faster the fit, we adopt as starting values for the velocities and the abundances those obtained with the two \texttt{cie} fit. After a first iteration we release the abundances and velocities of the two \texttt{pion} components in emission. The best-fit parameters are consistent, as number density and abundances, within uncertainties, with those obtained from a two-component \texttt{cie} model. The results are reported in Table~\ref{Table:Results spectral fits}. 
The model with two photoionised emission components yields $C_{\rm stat}/\mathrm{d.o.f.}=1320/1044$ (see Table~\ref{Table:Results spectral fits}). Adding a collisionally ionised \texttt{hot} absorber gives $1307/1042$ (\dcstat=13). With one blueshifted \texttt{xabs} absorber, the C-statistic decreases to 1296 (\dcstat=24). Adding a second blueshifted absorber lowers it by a further \dcstat=9, yielding $1287/1038$. Compared with the \texttt{hot} fit, the two-\texttt{xabs} fit has a C-statistic lower by 20 and four fewer degrees of freedom. Their $C_{\rm stat}/\mathrm{d.o.f.}$ ratios are close (1.24 and 1.25, respectively), and both absorption scenarios remain viable.

In a separate test (Appendix~\ref{sec:absorber_ufo_with_rest_absorber}), we repeated the \texttt{xabs} searches using the two-emitter configuration of Fig.~\ref{fig:2_PMF_absorber} and described in Sect. \ref{sec: density_emission lines}. This configuration includes a near-rest-frame \texttt{xabs} component to account for absorption in the O {\scriptsize VII} resonance line. The slower blueshifted candidate was recovered with \dcstat$\simeq24$ relative to this baseline (Fig.~\ref{fig:xabs_emission_density_correct_1xabs}), essentially the same improvement as in the main fits (Table~\ref{Table:Results spectral fits}). Adding the faster candidate yielded a further \dcstat $\simeq$15 in this test (Fig.~\ref{fig:xabs_emission_density_correct_2xabs}), compared with \dcstat $\simeq$ 9 in the main configuration. The faster candidate remains tentative.

We tested whether the absorption features could be produced by the ISM. We fitted the spectrum with, in turn, \texttt{amol} and an additional \texttt{hot} ISM absorbing component, following \citet{Pinto_2013}. In both cases, the improvement over the continuum model is modest (\dcstat$\sim10$), and neither configuration fully reproduces the broad absorption complex. In their sample of nine Galactic low-mass X-ray binaries (LMXBs), \citet{Pinto_2013} detected clear dust- and ISM-induced structures in the O K and Fe L edges for cold-gas hydrogen columns of $N_{\rm H}\sim(1$--$5)\times10^{21}\,{\rm cm^{-2}}$. For comparable elemental abundances and dust depletion, the lower column density along our line of sight would generally produce weaker neutral-gas and dust absorption. However, this comparison alone does not exclude an interstellar contribution, particularly from ionised gas. These tests therefore disfavour the specific ISM configurations considered here, but do not establish a unique physical origin for the absorption features. They refer to fits performed against the continuum-only baseline.

\subsection{Comparison with other sources}
A useful comparison can be made with another super-Eddington outburst presented by the Be/XRB pULX Swift J0243.6+6124 in 2017. \cite{Eijnden_2019} performed a high-resolution X-ray spectroscopy of the Chandra data, modelling the data with a spline continuum model. The description of the emission spectral lines was performed with the \texttt{photemis} and \texttt{bapec} in a photo-ionised gas and shock-ionised gas scenarios with the \texttt{XSPEC} software, respectively, assuming Solar abundances. The \texttt{photemis} provided a better description of the emission spectral lines with a log $\xi \sim 2.77 \pm 0.05 $ [erg /s cm]  with a turbulent velocity of $(2.2\pm 0.2) \times 10^{2}\rm \ km/s$ with an improvement of \dcstat $\sim 150 $, while the collisionally ionised model \texttt{bapec} gave a \dcstat $\sim$ 47. Moreover, the authors did not find any evidence of red- or blue-shifted lines. These results suggest a photo-ionised rest-frame emitting gas and are very similar to those found in this paper for the RX J0209.6 source.
For the absorption lines, instead, \citealt{Eijnden_2019} calculated, in Swift J0243.6+6124, the value of the blueshift expected for every combination of an observed absorption line and an observed higher-wavelength emission line. This value corresponds to $\sim$ -0.2c in the case of the outflow scenario involving the 8.4  \AA  $ \ $ line (Mg {\scriptsize XII} Ly$\alpha$), 9.50  \AA $ \ $ line (Ne {\scriptsize X} Ly$\delta$), and 12.15  \AA $ \ $ line (Ne {\scriptsize X} Ly$\alpha$), and providing an improvement of \dcstat $\sim$ 58 with respect to the continuum modelling. However, some questions arise about the physical interpretation of the absorption lines which might be considered, instead, all rest-frame and associated to Fe, Mg and Ca (\dcstat $\sim $ 142).

A0538--66 provides a close comparison with the source analysed in this paper, as both are transient Be/X-ray binaries in the Magellanic Clouds observed during super-Eddington outbursts. In the RGS spectra of A0538--66, \citet{Ducci_26} reported photoionised emission close to the systemic velocity and blueshifted absorption consistent with an outflow at $v_{\rm LOS}\simeq-0.12c$. This velocity is similar to that inferred for the slower candidate in RX J0209.6-7427.

NGC~300~ULX-1 provides another comparison with a pulsating ULX showing spectroscopic evidence for a relativistic outflow. In this case, \citealt{Kosec_2018b} reported no statistically significant improvement (\dcstat $\sim$ 14) when modelling the emission feature with the \texttt{photemis} and \texttt{cie} components, compared to a continuum model consisting of a double thermal component plus a power law (\texttt{bb+mbb+pow}). Moreover, they detected (with a significance greater than $3\sigma$) a rapidly variable (10s of ks), clumpy, ultrafast outflow at -0.22c, comparable to those observed in other ULXs, with a log $\xi\sim$  3.9, closer to the values typically measured in Active Galactic Nuclei (AGNs). This means that most of the absorption features are located at energies greater than 2 keV. In RX J0209.6, the initial searches for photoionised absorption identify a candidate outflow at $v_{\rm LOS}\sim-0.1c$, yielding an improvement of \dcstat$=58$ relative to the continuum-only model.
Adding a second absorber at $v_{\rm LOS}\sim-0.2c$ provides a further improvement of \dcstat$\simeq30$. Accounting for the emission lines and including the O {\scriptsize VII} resonance near-rest-frame absorber, the slower candidate outflow is recovered in the grid search, with \dcstat$\simeq24$ relative to this baseline (see Sect. \ref{subsec:Absorbing plasma in photo-ionisation equilibrium} and Fig. \ref{fig:xabs_emission_density_correct_1xabs}). Adding the faster component yields a further improvement of \dcstat$\simeq15$, and the evidence for this component remains tentative. However, the ambiguous line identifications, weak column-density constraints, and viable collisionally ionised absorption alternative prevent a conclusive identification of relativistic outflows. The luminosities and the column densities of the emitting plasma models, i.e. \texttt{cie/pion}, are much smaller than those in ULXs. This may be related to the shorter duration of the outburst which might correspond to just a mildly super-Eddington accretion disc in RX J0209.6 and, therefore, a lower mass loss. The volume and column densities estimated for the \texttt{pion} component are indeed comparable with those measured for Galactic XRBs (see, e.g., \citealt{Psaradaki_2018}). 

\section{Conclusion}
\label{sec:conclusion}
In this work, we performed an X-ray spectral analysis to search for a relativistic outflow during the super-Eddington outburst of RX J0209.6-7427, as expected in  super-Eddington accreting X-ray binaries. Two thermal components describe the emission of the inner and outer parts of the accretion disc, while a Comptonisation component takes into account the emission of the accretion column. This model provides a good fit for the continuum modelling, though spectral lines are present, both in emission and in absorption. The spectral modelling suggests the presence of a multiphase photoionised plasma approximately at rest with respect to the source, responsible for the observed emission features. The absorption features can be reproduced either by photoionised outflows moving towards the observer at approximately $0.1c$ and $0.2c$, the latter being more tentative, or by an alternative collisionally ionised absorption model, which provides a comparably acceptable fit. However, the ambiguities in the identification of the absorption lines and the relatively weak constraints on the absorber column densities prevent us from conclusively establishing the presence of relativistic outflows. Therefore, both interpretations remain viable. If associated with ultrafast outflows, the inferred column densities appear lower than those reported for some persistent ULXs (\citealt{Pinto_2017,Pinto_2020b}), potentially reflecting differences in the accretion-flow properties or the shorter duration of the super-Eddington phase in RX J0209.6-7427. Future high-resolution X-ray observations will be necessary to distinguish between these scenarios and determine the physical origin of the absorption features.

\section*{Data availability}
All datasets and software employed in this study are publicly accessible via ESA’s XMM-\textit{Newton} Science Archive (XSA, \url{https://www.cosmos.esa.int/web/XMM-Newton/xsa}) and NASA’s HEASARC archive (\url{https://heasarc.gsfc.nasa.gov/}). The spectral analysis codes and automated scanning routines developed for this work are also openly available on GitHub (\url{https://github.com/ciropinto1982}).

\begin{acknowledgements}
This work is based on the observation obtained with XMM-\textit{Newton}, an ESA science mission funded by the ESA Member States and the USA (NASA). FB and CP acknowledge support from PRIN MUR 2022 SEAWIND 2022Y2T94C funded by European Union - NextGenerationEU Mission 4 Component 1 CUP C53D23001330006, INAF Large Grant 2023 BLOSSOM O.F. 1.05.23.01.13 and INAF Large Grant 2024 "Timing the UltraLuminous X-ray Pulsars (TULiP)". T.D.S. acknowledges support from PRIN-INAF 2019 with the project “Probing the geometry of accretion: from theory to observations” (PI: Belloni). PK acknowledges support from NASA Hubble Fellowship grant HST-HF2-51534.001-A awarded by the Space Telescope Science Institute and from NASA grants 80NSSC25K7317 and 80NSSC25K7533.
\end{acknowledgements}

\bibliographystyle{aa}
\bibliography{bibliography}

@ARTICLE{Ducci_26,
       author = {{Ducci}, Lorenzo and {Mereghetti}, Sandro and {Pinto}, Ciro and {Bachetti}, Matteo and {Nakajima}, Motoki and {Sharma}, Rahul and {Covino}, Stefano and {Rigoselli}, Michela and {Tresoldi}, Caterina and {Mihara}, Tatehiro and et al.},
        title = "{A0538-66: A transient ultraluminous X-ray source with a relativistic outflow, 69 ms pulsations, and anomalous optical activity}",
      journal = {\aap},
         year = 2026,
        month = nov,
       volume = {715},
          eid = {A170},
        pages = {A170},
          doi = {10.1051/0004-6361/202663088},
archivePrefix = {arXiv},
       eprint = {2610.17982},
 primaryClass = {astro-ph.HE},
       adsurl = {https://scixplorer.org/abs/2026A&A...715A.170D}
}

@ARTICLE{Vasilopoulos_2020,
       author = {{Vasilopoulos}, G. and {Lander}, S.~K. and {Koliopanos}, F. and {Bailyn}, C.~D.},
        title = "{M51 ULX-7: superorbital periodicity and constraints on the neutron star magnetic field}",
      journal = {\mnras},
         year = 2020,
        month = feb,
       volume = {491},
       number = {4},
        pages = {4949-4959},
          doi = {10.1093/mnras/stz3298},
archivePrefix = {arXiv},
       eprint = {1911.09670},
 primaryClass = {astro-ph.HE},
       adsurl = {https://ui.adsabs.harvard.edu/abs/2020MNRAS.491.4949V}
}

@ARTICLE{Iwakiri_2019,
       author = {{Iwakiri}, W. and {Wolff}, M.~T. and {Vasilopoulos}, G. and {Gendreau}, K. and {Arzoumanian}, Z. and {Markwardt}, C. and {Strohmayer}, T.~E. and {Ray}, P.~S. and {Altamirano}, D. and {Mihara}, T. and {Negoro}, H. and {Guillot}, S. and {Jaisawal}, G.~K.},
        title = "{NICER detection of 9.29 s pulsations from RX J0209.6-7427}",
      journal = {The Astronomer's Telegram},
         year = 2019,
        month = nov,
       volume = {13309},
        pages = {1},
       adsurl = {https://ui.adsabs.harvard.edu/abs/2019ATel13309....1I}
}

@ARTICLE{West_2017,
       author = {{West}, Brent F. and {Wolfram}, Kenneth D. and {Becker}, Peter A.},
        title = "{A New Two-fluid Radiation-hydrodynamical Model for X-Ray Pulsar Accretion Columns}",
      journal = {\apj},
         year = 2017,
        month = feb,
       volume = {835},
       number = {2},
          eid = {129},
        pages = {129},
          doi = {10.3847/1538-4357/835/2/129},
archivePrefix = {arXiv},
       eprint = {1612.02411},
 primaryClass = {astro-ph.HE},
       adsurl = {https://ui.adsabs.harvard.edu/abs/2017ApJ...835..129W}
}

@ARTICLE{Bachetti_2020,
       author = {{Bachetti}, Matteo and {Maccarone}, Thomas J. and {Brightman}, Murray and {Brumback}, McKinley C. and {F{\"u}rst}, Felix and {Harrison}, Fiona A. and {Heida}, Marianne and {Israel}, Gian Luca and {Middleton}, Matthew J. and {Tomsick}, John A. and {Webb}, Natalie A. and {Walton}, Dominic J.},
        title = "{All at Once: Transient Pulsations, Spin-down, and a Glitch from the Pulsating Ultraluminous X-Ray Source M82 X-2}",
      journal = {\apj},
         year = 2020,
        month = mar,
       volume = {891},
       number = {1},
          eid = {44},
        pages = {44},
          doi = {10.3847/1538-4357/ab6d00},
archivePrefix = {arXiv},
       eprint = {1905.06423},
 primaryClass = {astro-ph.HE},
       adsurl = {https://ui.adsabs.harvard.edu/abs/2020ApJ...891...44B}
}

@ARTICLE{Walton_2018b,
       author = {{Walton}, D.~J. and {Bachetti}, M. and {F{\"u}rst}, F. and {Barret}, D. and {Brightman}, M. and {Fabian}, A.~C. and {Grefenstette}, B.~W. and {Harrison}, F.~A. and {Heida}, M. and {Kennea}, J. and {Kosec}, P. and {Lau}, R.~M. and {Madsen}, K.~K. and {Middleton}, M.~J. and {Pinto}, C. and {Steiner}, J.~F. and {Webb}, N.},
        title = "{A Potential Cyclotron Resonant Scattering Feature in the Ultraluminous X-Ray Source Pulsar NGC 300 ULX1 Seen by NuSTAR and XMM-Newton}",
      journal = {\apjl},
         year = 2018,
        month = apr,
       volume = {857},
       number = {1},
          eid = {L3},
        pages = {L3},
          doi = {10.3847/2041-8213/aabadc},
archivePrefix = {arXiv},
       eprint = {1803.07571},
 primaryClass = {astro-ph.HE},
       adsurl = {https://ui.adsabs.harvard.edu/abs/2018ApJ...857L...3W}
}

@ARTICLE{Vasilopoulos_2019,
       author = {{Vasilopoulos}, G. and {Petropoulou}, M. and {Koliopanos}, F. and {Ray}, P.~S. and {Bailyn}, C.~B. and {Haberl}, F. and {Gendreau}, K.},
        title = "{NGC 300 ULX1: spin evolution, super-Eddington accretion, and outflows}",
      journal = {\mnras},
         year = 2019,
        month = oct,
       volume = {488},
       number = {4},
        pages = {5225-5231},
          doi = {10.1093/mnras/stz2045},
archivePrefix = {arXiv},
       eprint = {1905.03740},
 primaryClass = {astro-ph.HE},
       adsurl = {https://ui.adsabs.harvard.edu/abs/2019MNRAS.488.5225V} }

@ARTICLE{Cash_1979,
       author = {{Cash}, W.},
        title = "{Parameter estimation in astronomy through application of the likelihood ratio.}",
      journal = {\apj},
         year = 1979,
        month = mar,
       volume = {228},
        pages = {939-947},
          doi = {10.1086/156922},
       adsurl = {https://ui.adsabs.harvard.edu/abs/1979ApJ...228..939C}
}

@ARTICLE{Reig_2011,
       author = {{Reig}, Pablo},
        title = "{Be/X-ray binaries}",
      journal = {\apss},
         year = 2011,
        month = mar,
       volume = {332},
       number = {1},
        pages = {1-29},
          doi = {10.1007/s10509-010-0575-8},
archivePrefix = {arXiv},
       eprint = {1101.5036},
 primaryClass = {astro-ph.HE},
       adsurl = {https://ui.adsabs.harvard.edu/abs/2011Ap&SS.332....1R}
}

@ARTICLE{Eijnden_2019,
       author = {{van den Eijnden}, J. and {Degenaar}, N. and {Schulz}, N.~S. and {Nowak}, M.~A. and {Wijnands}, R. and {Russell}, T.~D. and {Hern{\'a}ndez Santisteban}, J.~V. and {Bahramian}, A. and {Maccarone}, T.~J. and {Kennea}, J.~A. and {Heinke}, C.~O.},
        title = "{Chandra reveals a possible ultrafast outflow in the super-Eddington Be/X-ray binary Swift J0243.6+6124}",
      journal = {\mnras},
         year = 2019,
        month = aug,
       volume = {487},
       number = {3},
        pages = {4355-4371},
          doi = {10.1093/mnras/stz1548},
archivePrefix = {arXiv},
       eprint = {1906.01597},
 primaryClass = {astro-ph.HE},
       adsurl = {https://ui.adsabs.harvard.edu/abs/2019MNRAS.487.4355V}
}

@ARTICLE{King_2019,
       author = {{King}, Andrew and {Lasota}, Jean-Pierre},
        title = "{No magnetars in ULXs}",
      journal = {\mnras},
         year = 2019,
        month = may,
       volume = {485},
       number = {3},
        pages = {3588-3594},
          doi = {10.1093/mnras/stz720},
archivePrefix = {arXiv},
       eprint = {1903.03624},
 primaryClass = {astro-ph.HE},
       adsurl = {https://ui.adsabs.harvard.edu/abs/2019MNRAS.485.3588K}
}

@ARTICLE{Mushtukov_2015,
       author = {{Mushtukov}, Alexander A. and {Suleimanov}, Valery F. and {Tsygankov}, Sergey S. and {Poutanen}, Juri},
        title = "{The critical accretion luminosity for magnetized neutron stars}",
      journal = {\mnras},
         year = 2015,
        month = feb,
       volume = {447},
       number = {2},
        pages = {1847-1856},
          doi = {10.1093/mnras/stu2484},
archivePrefix = {arXiv},
       eprint = {1409.6457},
 primaryClass = {astro-ph.HE},
       adsurl = {https://ui.adsabs.harvard.edu/abs/2015MNRAS.447.1847M}
}

@ARTICLE{Barra_2024,
       author = {{Barra}, F. and {Pinto}, C. and {Middleton}, M. and {Di Salvo}, T. and {Walton}, D.~J. and {G{\'u}rpide}, A. and {Roberts}, T.~P.},
        title = "{Examining the nature of the ultraluminous X-ray source Holmberg II X-1}",
      journal = {\aap},
         year = 2024,
        month = feb,
       volume = {682},
          eid = {A94},
        pages = {A94},
          doi = {10.1051/0004-6361/202348471},
archivePrefix = {arXiv},
       eprint = {2311.16243},
 primaryClass = {astro-ph.HE},
       adsurl = {https://ui.adsabs.harvard.edu/abs/2024A&A...682A..94B}
}

@ARTICLE{Bachetti_2014,
   author = {{Bachetti}, M. and {Harrison}, F.~A. and {Walton}, D.~J. and 
	{Grefenstette}, B.~W. and {Chakrabarty}, D. and others},
    title = "{An ultraluminous X-ray source powered by an accreting neutron star}",
  journal = {\nat},
archivePrefix = "arXiv",
   eprint = {1410.3590},
 primaryClass = "astro-ph.HE",
     year = 2014,
    month = oct,
   volume = 514,
    pages = {202-204},
      doi = {10.1038/nature13791},
   adsurl = {http://adsabs.harvard.edu/abs/2014Natur.514..202B}
}

@ARTICLE{Caserta_2026,
       author = {{Caserta}, S. and {Pinto}, C. and {Di Salvo}, T. and {Pintore}, F. and {Kosec}, P. and {Barra}, F. and {Walton}, D.~J. and {D'A{\`\i}}, A. and {Del Santo}, M. and {G{\'u}rpide}, A. and {Fabian}, A. and {Wolter}, A.},
        title = "{Unveiling the biconical geometry of the outflow in the ultraluminous X-ray source NGC 5204 X-1}",
      journal = {\aap},
         year = 2026,
        month = apr,
       volume = {708},
          eid = {A347},
        pages = {A347},
          doi = {10.1051/0004-6361/202558687},
archivePrefix = {arXiv},
       eprint = {2603.11922},
 primaryClass = {astro-ph.HE},
       adsurl = {https://ui.adsabs.harvard.edu/abs/2026A&A...708A.347C}
}

@ARTICLE{Barra_2025,
       author = {{Barra}, F. and {Barret}, D. and {Pinto}, C. and {Di Salvo}, T. and {Weinberg}, N. and {Guichandut}, S.},
        title = "{Line detections in photospheric radius expansion bursts from 4U 1820-303: Confirmation of previous detections and their temporal evolution}",
      journal = {\aap},
         year = 2025,
        month = feb,
       volume = {694},
          eid = {A266},
        pages = {A266},
          doi = {10.1051/0004-6361/202452878},
archivePrefix = {arXiv},
       eprint = {2501.01488},
 primaryClass = {astro-ph.HE},
       adsurl = {https://ui.adsabs.harvard.edu/abs/2025A&A...694A.266B}
}

@ARTICLE{Tarter_1969,
       author = {{Tarter}, C. Bruce and {Tucker}, Wallace H. and {Salpeter}, Edwin E.},
        title = "{The Interaction of X-Ray Sources with Optically Thin Environments}",
      journal = {\apj},
         year = 1969,
        month = jun,
       volume = {156},
        pages = {943},
          doi = {10.1086/150026},
       adsurl = {https://ui.adsabs.harvard.edu/abs/1969ApJ...156..943T}
}

@ARTICLE{Krolik_1981,
       author = {{Krolik}, J.~H. and {McKee}, C.~F. and {Tarter}, C.~B.},
        title = "{Two-phase models of quasar emission line regions.}",
      journal = {\apj},
         year = 1981,
        month = oct,
       volume = {249},
        pages = {422-442},
          doi = {10.1086/159303},
       adsurl = {https://ui.adsabs.harvard.edu/abs/1981ApJ...249..422K}
}

@ARTICLE{Israel_2016,
       author = {{Israel}, G.~L. and {Esposito}, P. and {Rodr{\'\i}guez Castillo}, G.~A. and {Sidoli}, L.},
        title = "{The Chandra ACIS Timing Survey Project: glimpsing a sample of faint X-ray pulsators}",
      journal = {\mnras},
         year = 2016,
        month = nov,
       volume = {462},
       number = {4},
        pages = {4371-4385},
          doi = {10.1093/mnras/stw1897},
archivePrefix = {arXiv},
       eprint = {1608.00077},
 primaryClass = {astro-ph.HE},
       adsurl = {https://ui.adsabs.harvard.edu/abs/2016MNRAS.462.4371I}
}

@ARTICLE{Kaaret_2017,
       author = {{Kaaret}, Philip and {Feng}, Hua and {Roberts}, Timothy P.},
        title = "{Ultraluminous X-Ray Sources}",
      journal = {\araa},
         year = "2017",
        month = "Aug",
       volume = {55},
       number = {1},
        pages = {303-341},
          doi = {10.1146/annurev-astro-091916-055259},
archivePrefix = {arXiv},
       eprint = {1703.10728},
 primaryClass = {astro-ph.HE},
       adsurl = {https://ui.adsabs.harvard.edu/abs/2017ARA&A..55..303K}
}

@ARTICLE{Kaastra_2016,
       author = {{Kaastra}, J.~S. and {Bleeker}, J.~A.~M.},
        title = "{Optimal binning of X-ray spectra and response matrix design}",
      journal = {\aap},
         year = 2016,
        month = mar,
       volume = {587},
          eid = {A151},
        pages = {A151},
          doi = {10.1051/0004-6361/201527395},
archivePrefix = {arXiv},
       eprint = {1601.05309},
 primaryClass = {astro-ph.IM},
       adsurl = {https://ui.adsabs.harvard.edu/abs/2016A&A...587A.151K}
}

@misc{kaastra2023_spex,
  author       = {Kaastra, J. S. and
                  Raassen, A. J. J. and
                  de Plaa, J. and
                  Gu, Liyi},
  title        = {SPEX X-ray spectral fitting package},
  month        = march,
  year         = 2023,
  publisher    = {Zenodo},
  doi          = {10.5281/zenodo.7777289},
  url          = {https://doi.org/10.5281/zenodo.7777289}
}

@ARTICLE{Kosec_2021,
       author = {{Kosec}, P. and {Pinto}, C. and {Reynolds}, C.~S. and {Guainazzi}, M. and {Kara}, E. and {Walton}, D.~J. and {Fabian}, A.~C. and {Parker}, M.~L. and {Valtchanov}, I.},
        title = "{Ionized emission and absorption in a large sample of ultraluminous X-ray sources}",
      journal = {\mnras},
         year = 2021,
        month = dec,
       volume = {508},
       number = {3},
        pages = {3569-3588},
          doi = {10.1093/mnras/stab2856},
archivePrefix = {arXiv},
       eprint = {2109.14683},
 primaryClass = {astro-ph.HE},
       adsurl = {https://ui.adsabs.harvard.edu/abs/2021MNRAS.508.3569K}
}

@ARTICLE{Kosec_2018a,
   author = {{Kosec}, P. and {Pinto}, C. and {Fabian}, A.~C. and {Walton}, D.~J.
	},
    title = "{Searching for outflows in ultraluminous X-ray sources through high-resolution X-ray spectroscopy}",
  journal = {\mnras},
archivePrefix = "arXiv",
   eprint = {1710.06438},
 primaryClass = "astro-ph.HE",
     year = {{2018}},
    month = feb,
   volume = 473,
    pages = {5680-5697},
      doi = {10.1093/mnras/stx2695},
   adsurl = {http://adsabs.harvard.edu/abs/2018MNRAS.473.5680K}
}

@ARTICLE{Kosec_2018b,
   author = {{Kosec}, P. and {Pinto}, C. and {Walton}, D.~J. and others},
    title = "{Evidence for a variable Ultrafast Outflow in the newly discovered Ultraluminous Pulsar NGC 300 ULX-1}",
  journal = {\mnras},
archivePrefix = "arXiv",
   eprint = {1803.02367},
 primaryClass = "astro-ph.HE",
     year = {{2018}},
    month = sep,
   volume = 479,
    pages = {3978-3986},
      doi = {10.1093/mnras/sty1626},
   adsurl = {http://adsabs.harvard.edu/abs/2018MNRAS.479.3978K}
}

@ARTICLE{Lodders2009,
   author = {{Lodders}, K. and {Palme}, H.},
    title = "{Solar System Elemental Abundances in 2009}",
  journal = {Meteoritics and Planetary Science Supplement},
     year = 2009,
    month = sep,
   volume = 72,
    pages = {5154-+},
   adsurl = {http://adsabs.harvard.edu/abs/2009M%26PSA..72.5154L}
}

@ARTICLE{Pinto_2017,
   author = {{Pinto}, C. and {Alston}, W. and {Soria}, R. and {Middleton}, M.~J. and 
	{Walton}, D.~J. and others},
    title = "{From ultraluminous X-ray sources to ultraluminous supersoft sources: NGC 55 ULX, the missing link}",
  journal = {\mnras},
archivePrefix = "arXiv",
   eprint = {1612.05569},
 primaryClass = "astro-ph.HE",
     year = 2017,
    month = jul,
   volume = 468,
    pages = {2865-2883},
      doi = {10.1093/mnras/stx641},
   adsurl = {http://adsabs.harvard.edu/abs/2017MNRAS.468.2865P}
}

@ARTICLE{Pinto_2013,
       author = {{Pinto}, C. and {Kaastra}, J.~S. and {Costantini}, E. and {de Vries}, C.},
        title = "{Interstellar medium composition through X-ray spectroscopy of low-mass X-ray binaries}",
      journal = {\aap},
         year = 2013,
        month = mar,
       volume = {551},
          eid = {A25},
        pages = {A25},
          doi = {10.1051/0004-6361/201220481},
archivePrefix = {arXiv},
       eprint = {1301.1612},
 primaryClass = {astro-ph.GA},
       adsurl = {https://ui.adsabs.harvard.edu/abs/2013A&A...551A..25P}
}

@ARTICLE{Pinto_2016,
   author = {{Pinto}, C. and {Middleton}, M.~J. and {Fabian}, A.~C.},
    title = "{Resolved atomic lines reveal outflows in two ultraluminous X-ray sources}",
  journal = {Nature},
archivePrefix = "arXiv",
   eprint = {1604.08593},
 primaryClass = "astro-ph.HE",
     year = 2016,
    month = may,
   volume = 533,
    pages = {64-67},
      doi = {10.1038/nature17417},
   adsurl = {http://adsabs.harvard.edu/abs/2016Natur.533...64P}
}

@ARTICLE{Pinto_2020a,
       author = {{Pinto}, C. and {Mehdipour}, M. and {Walton}, D.~J. and others},
        title = "{Thermal stability of winds driven by radiation pressure in super-Eddington accretion discs}",
      journal = {\mnras},
         year = 2020,
        month = feb,
       volume = {491},
       number = {4},
        pages = {5702-5716},
          doi = {},
archivePrefix = {arXiv},
       eprint = {},
 primaryClass = {},
       adsurl = {https://ui.adsabs.harvard.edu/abs/2020MNRAS.491.5702P}
}

@ARTICLE{Pinto_2020b,
       author = {{Pinto}, C. and {Walton}, D.~J. and {Kara}, E. and {Parker}, M.~L. and {Soria}, R. and others},
        title = "{XMM-Newton campaign on ultraluminous X-ray source NGC 1313 X-1: wind versus state variability}",
      journal = {\mnras},
         year = 2020,
        month = mar,
       volume = {492},
       number = {4},
        pages = {4646-4665},
          doi = {10.1093/mnras/staa118},
archivePrefix = {arXiv},
       eprint = {1911.09568},
 primaryClass = {astro-ph.HE},
       adsurl = {https://ui.adsabs.harvard.edu/abs/2020MNRAS.492.4646P}
}

@ARTICLE{Pinto_2023,
       author = {{Pinto}, Ciro and {Walton}, Dominic J.},
        title = "{Ultra-luminous X-ray sources: extreme accretion and feedback}",
      journal = {arXiv e-prints},
         year = 2023,
        month = jan,
          eid = {arXiv:2302.00006},
        pages = {arXiv:2302.00006},
          doi = {10.48550/arXiv.2302.00006},
archivePrefix = {arXiv},
       eprint = {2302.00006},
 primaryClass = {astro-ph.HE},
       adsurl = {https://ui.adsabs.harvard.edu/abs/2023arXiv230200006P}
}

@ARTICLE{Pinto_2025,
       author = {{Pinto}, C. and {Caserta}, S. and {Barra}, F. and {Xu}, Y. and {Barret}, D. and {Kosec}, P. and {La Palombara}, N. and {Marino}, A. and {Pintore}, F. and {Riggio}, A. and et al.},
        title = "{XMM-Newton multi-year campaign on NGC 55 ULX-1: Resolving the wind and its variability with RGS}",
      journal = {\aap},
         year = 2026,
        month = feb,
       volume = {706},
          eid = {A266},
        pages = {A266},
          doi = {10.1051/0004-6361/202556631},
archivePrefix = {arXiv},
       eprint = {2512.16522},
 primaryClass = {astro-ph.HE},
       adsurl = {https://scixplorer.org/abs/2026A&A...706A.266P}
}

@ARTICLE{Psaradaki_2018,
       author = {{Psaradaki}, I. and {Costantini}, E. and {Mehdipour}, M. and {D{\'\i}az Trigo}, M.},
        title = "{Modelling the disc atmosphere of the low mass X-ray binary EXO 0748-676}",
      journal = {\aap},
         year = 2018,
        month = dec,
       volume = {620},
          eid = {A129},
        pages = {A129},
          doi = {10.1051/0004-6361/201834000},
archivePrefix = {arXiv},
       eprint = {1809.06864},
 primaryClass = {astro-ph.HE},
       adsurl = {https://ui.adsabs.harvard.edu/abs/2018A&A...620A.129P}
}

@ARTICLE{Rodriguez_2020,
       author = {{Rodr{\'\i}guez Castillo}, G.~A. and {Israel}, G.~L. and {Belfiore}, A. and {Bernardini}, F. and {Esposito}, P. and {Pintore}, F. and {De Luca}, A. and {Papitto}, A. and {Stella}, L. and {Tiengo}, A. and {Zampieri}, L. and {Bachetti}, M. and {Brightman}, M. and {Casella}, P. and {D'Agostino}, D. and {Dall'Osso}, S. and {Earnshaw}, H.~P. and {F{\"u}rst}, F. and {Haberl}, F. and {Harrison}, F.~A. and {Mapelli}, M. and {Marelli}, M. and {Middleton}, M. and {Pinto}, C. and {Roberts}, T.~P. and {Salvaterra}, R. and {Turolla}, R. and {Walton}, D.~J. and {Wolter}, A.},
        title = "{Discovery of a 2.8 s Pulsar in a 2 Day Orbit High-mass X-Ray Binary Powering the Ultraluminous X-Ray Source ULX-7 in M51}",
      journal = {\apj},
         year = 2020,
        month = may,
       volume = {895},
       number = {1},
          eid = {60},
        pages = {60},
          doi = {10.3847/1538-4357/ab8a44},
archivePrefix = {arXiv},
       eprint = {1906.04791},
 primaryClass = {astro-ph.HE},
       adsurl = {https://ui.adsabs.harvard.edu/abs/2020ApJ...895...60R}
}

@ARTICLE{Pourquet_2000,
       author = {{Porquet}, D. and {Dubau}, J.},
        title = "{X-ray photoionized plasma diagnostics with helium-like ions. Application to warm absorber-emitter in active galactic nuclei}",
      journal = {\aaps},
         year = 2000,
        month = may,
       volume = {143},
        pages = {495-514},
          doi = {10.1051/aas:2000192},
archivePrefix = {arXiv},
       eprint = {astro-ph/0002319},
 primaryClass = {astro-ph},
       adsurl = {https://ui.adsabs.harvard.edu/abs/2000A&AS..143..495P}
}

\begin{appendix}
\begin{sidewaystable*}[p]
\section{Best-fit spectral parameters}
\centering
\caption{Best-fit results of the spectral models.}
\label{Table:Results spectral fits}

\renewcommand{\arraystretch}{1.2}
\scriptsize
\setlength{\tabcolsep}{3pt}

\resizebox{0.97\textheight}{!}{%
\begin{tabular}{*{14}{c}}
    \toprule
   {{Parameter /}} & {{Continuum}} &  {{PION}} & {{PION (abs.)}} & {{XABS}} & {{CIE}} & {{PION + XABS}}  & {{2 XABS}} & {{PION + 2 XABS}} & {{2 CIE}} & {{2 PION}} & {{2 PION + HOT}} & {{2 PION + XABS}} & {{2 PION + 2 XABS}} \\
    {{component}} & {{model}} & {{model}} & {{model}} & {{model}} & {{model}} & {{model}} & {{model}}& {{model}} & {{model}} & {{model}} & {{model}} & {{model}} & {{model}} \\
    
    \midrule
$L_{X,\rm mbb}$           & 0.43 $\pm$ 0.02                 & --   & --  & --   & --  & --  & --  & --  & -- & --  & -- & -- & -- \\
$L_{X,\rm mbb_2}$         & 0.72 $\pm$ 0.13                 & --   & --  & --   & --  & --  & --  & --  & -- & --  & -- & -- & -- \\
kT$_{\rm mbb}$            & 0.259 $\pm$ 0.005               & --   & --  & --   & --  & --  & --  & --  & -- &--   &-- & -- & -- \\
kT$_{\rm mbb,2}$          & 1.0 $\pm$  0.2                  & --   & --  & --   & --  & --  & --  & --  & -- & --  & -- & -- & -- \\
 $L_{X,\rm comt}$         & 3.2 $\pm$  0.2                  & --   & --  & --   & --  & --  & --  & --  & -- &--   &-- & -- & -- \\
 $L_{\rm Bol,\rm Tot}$    & 9.5 $\pm$  0.2                  & --   & --  & --   & --  & --  & --  & --  & -- & --  & -- & -- & -- \\
   k$\rm{T_e}$            & 3.6 $\pm$  0.1                  & --   & --  & --   & --  & --  & --  & --  & -- & --  & -- & -- & -- \\
  k$\rm{T_{seed}}$        & 1.0 (coupled)                   & --   & --  & --   & --  & --  & --  & --  & -- & --  & -- & -- & -- \\
    $\tau$                & $9.98_{-0.38}^{+0.07}$    & --   & --  & --   & --  & --  & --  & --  & -- & --  & -- & -- & -- \\

$N_{\rm H}$               & 0.58 $\pm$ 0.01                 & --   & --  & --   & --  & --  & --  & --  & -- & -- & -- & -- & -- \\
\hline
$\rm L_{pion}$            & --                              & 0.022 $\pm$ 0.003 &  --  & --  &  -- & 0.021 $\pm$ 0.003 & -- & 0.020 $\pm$ 0.003  & -- & 0.029 $\pm$ 0.002                            & 0.030 $\pm$ 0.001 & 0.033 $\pm$ 0.005 & 0.032 $\pm$ 0.004 \\
$\rm N_{H,pion}$          & --                              & 3.5 $\pm$ 0.5     &  0.10 $\pm$ 0.01      & --  &  -- & 3.3 $\pm$ 0.5     & -- & 3.1 $\pm$ 0.5      & -- & 6.3 $\pm$ 0.5               & $6.27_{-0.3}^{+0.6}$ & 6.9 $\pm$ 1.0  & 6.8 $\pm$ 1.0 \\
Log $\xi_{\rm pion}$      & --                              &  2.62 $\pm$ 0.03  &   0.69 $\pm$ 0.07   & --  &  -- & 2.63 $\pm$ 0.04   & -- & 2.62 $\pm$ 0.04    & -- & 2.61 $\pm$ 0.02             & $2.61_{-0.03}^{+0.02}$ & 2.61 $\pm$ 0.03 & 2.63 $\pm$ 0.05 \\
$\rm v_{LOS,pion}$        & --                              & 100 $\pm$ 300     &  -30000 $\pm$ 300      & --  &  -- & 100 $\pm$ 300     & -- & 125 $\pm$ 300      & -- & -170 $\pm$ 180            & -170 $\pm$ 10 & -100 $\pm$ 200 & -100 $\pm$ 200 \\
$\rm v_{RMS,pion}$        & --                              & 1000 (fixed)      &  1000 (fixed)       & --  &  -- & 1000 (fixed)      & -- & 1000 (fixed)       & -- & $2000_{-400}^{+200}$   & 1600 $\pm$ 250 & 1900 $\pm$ 300 & 1800 $\pm$ 300 \\
$\rm L_{pion_2}$          & --                              & --                &  --                 & --  &  -- & --                & -- & --                 & -- & 0.010 $\pm$ 0.001           & 0.010 $\pm$ 0.001 & 0.005 $\pm$ 0.002  & 0.005 $\pm$ 0.002 \\
$\rm N_{H,pion_2}$        & --                              & --                &  --                 & --  &  -- & --                & -- &--                  & -- & 0.17 $\pm$ 0.02             & 0.17 $\pm$ 0.02 & 0.11 $\pm$ 0.04 & 0.11 $\pm$ 0.04 \\
Log $\xi_{\rm pion_2}$    & --                              &  --               &  --                & --  &  -- & --                & -- &--                  & -- & $1.16_{-0.10}^{+0.14}$  & $1.17_{-0.10}^{+0.13}$ & 1.3 $\pm$ 0.2  & 1.3 $\pm$ 0.2 \\
$\rm v_{LOS,pion_2}$      & --                              & --                &  --                 & --  &  -- & --                & -- & --                 & -- & -170 (coupled)              & -170 (coupled) & -100 (coupled)  & -100 (coupled) \\
$\rm v_{RMS,pion_2}$      & --                              & --                &  --                 & --  &  -- & --                & -- & --                 & -- & 2000 (coupled)              & 1600 (coupled) & 1900 (coupled) & 1800 (coupled) \\
\hline               
$\rm N_{H,xabs/hot}$          & -- & -- & --   & 0.09 $\pm$ 0.01       & --  & 0.07 $\pm$ 0.01    & 0.09 $\pm$ 0.01   & 0.07 $\pm$ 0.01     & -- & -- & 0.14 $\pm$ 0.02 & 0.04 $\pm$ 0.02 & 0.05 $\pm$ 0.02 \\
Log $\xi_{\rm xabs}$          & -- & -- & --   & 0.70 $\pm$ 0.07       & --  & 0.8 $\pm$  0.1     & 0.75 $\pm$ 0.07   & 0.83 $\pm$ 0.11         & -- & --  & -- & 0.87 $\pm$ 0.12 & 0.85 $\pm$ 0.10 \\
$\rm kT_{\rm hot}$            & -- & -- & --   & --                    & --  & --                 & --                & --          & --  & --  & 0.019 $\pm$ 0.002 & -- & -- \\
$\rm v_{LOS,xabs/hot}$        & -- & -- & --   & -30000 $\pm$ 300      & --  & -30000 $\pm$  400  & -30000 $\pm$ 300  & -30000 $\pm$ 400    & -- & --  & 0 (fixed) & -30000 $\pm$ 300 & -30000 $\pm$ 300 \\
$\rm v_{RMS,xabs/hot}$        & -- & -- & --   & 1000 (fixed)          & --  & 1000 (fixed)       & 1000 (fixed)      & 1000 (fixed)        & -- & --  & 100 (fixed) & 1000 (fixed) & 1000 (fixed) \\
$\rm N_{H,xabs_2}$            & -- & -- & --   & --                    & --  & --                 & 0.06 $\pm$ 0.01   & 0.04 $\pm$ 0.01     & -- & --   & -- & -- & 0.03 $\pm$ 0.02 \\
Log $\xi_{\rm xabs_2}$        & -- & -- & --   & --                    & --  & --                 & 0.62 $\pm$ 0.09   & 0.65 $\pm$ 0.12     & -- & --   & -- & -- & 0.6 $\pm$ 0.1 \\
$\rm v_{LOS,xabs_2}$          & -- & -- & --   & --                    & --  & --                 & -69000 $\pm$ 400  & -69000 $\pm$ 400     & -- & --   & -- & -- & -69000 $\pm$ 400 \\
$\rm v_{RMS,xabs_2}$          & -- & -- & --   & --                    & --  & --                 & 1000 (fixed)      & 1000 (fixed)        & -- & --   & -- & -- & 1000 (fixed) \\
\hline 
$\rm L_{CIE}$        & -- & --      &  --   & --        & 0.040 $\pm$ 0.005        & --       & -- & --         & 0.041 $\pm$  0.005 & --  & -- & -- & -- \\
$\rm Norm_{CIE}$     & -- & --      &  --   & --        & 25 $\pm$ 3               & --       & -- &--          & 29 $\pm$ 4         & --  & -- & -- & -- \\
$\rm kT_{\rm CIE}$   & -- & --      &  --   & --        & 1.25 $\pm$ 0.05          & --       & -- &--          & 1.07 $\pm$ 0.05    & --  & -- & -- & -- \\
$\rm v_{LOS,CIE}$    & -- & --      &  --   & --        & 3 $\pm$ 300              & --       & -- &--          & 10 $\pm$ 200       & --  & -- & -- & -- \\
$\rm v_{RMS,CIE}$    & -- & --      &  --   & --        & 1000 (fixed)             & --       & -- &--          & 1700 $\pm$ 300     & --  & -- & -- & -- \\
$\rm L_{CIE_2}$      & -- & --      &  --   & --        & --                       & --       & -- & --         & 0.010 $\pm$ 0.003  & --  & -- & -- & -- \\
$\rm Norm_{CIE_2}$   & -- & --      &  --   & --        & --                       & --       & -- &--          &  10 $\pm$ 2        & --  & -- & -- & -- \\
$\rm kT_{\rm CIE_2}$ & -- & --      &  --   & --        & --                       & --       & -- &--          & 0.16 $\pm$ 0.01    & --  & -- & -- & -- \\
$\rm v_{LOS,CIE_2}$  & -- & --      &  --   & --        & --                       & --       & -- &--          & 10 (coupled)       & --  & -- & -- & -- \\
$\rm v_{RMS,CIE_2}$  & -- & --      &  --   & --        & --                       & --       & -- &--          & 1700 (coupled)     & --  & -- & -- & -- \\
\hline
${\rm C_{stat}}$/d.o.f & 1559/1053  & 1405/1050  & 1501/1050  & 1501/1050 &  1434/1050  & 1368/1047 & 1473/1047 & 1353/1044 & 1348/1044 & 1320/1044  & 1307/1042 & 1296/1041 & 1287/1038 \\
\hline
\end{tabular}%
}
\tablefoot
    {The hydrogen column density, $N_{\rm H}$, of the \texttt{hot}, \texttt{pion} and \texttt{xabs} components are expressed in units of $10^{21}\ \rm cm^{-2}$.
     The X-ray and bolometric luminosities are expressed in $10^{38}$ erg/s unit in the 0.3-10 keV and 0.001-1000 keV energy band, respectively. The line of sight $\rm v_{LOS}$ and the turbulent velocity $\rm v_{RMS}$ are expressed in km/s. The temperatures kT are expressed in keV unit. $\tau$ is the optical depth of the Comptonisation component. The continuum parameters, with the exception of the normalisations, were fixed to the best-fit continuum values during the fits with the optically-thin plasma models. $\rm Norm_{CIE}$ (i.e. emission measure) are in $10^{58}$\,cm$^{-3}$ units.}
\end{sidewaystable*}

\clearpage
\onecolumn
\section{Collisional and photoionised emission grids} 
\label{appendix-number_density}

\begin{figure*}[h!]

    \centering
    \includegraphics[width=.45\linewidth]{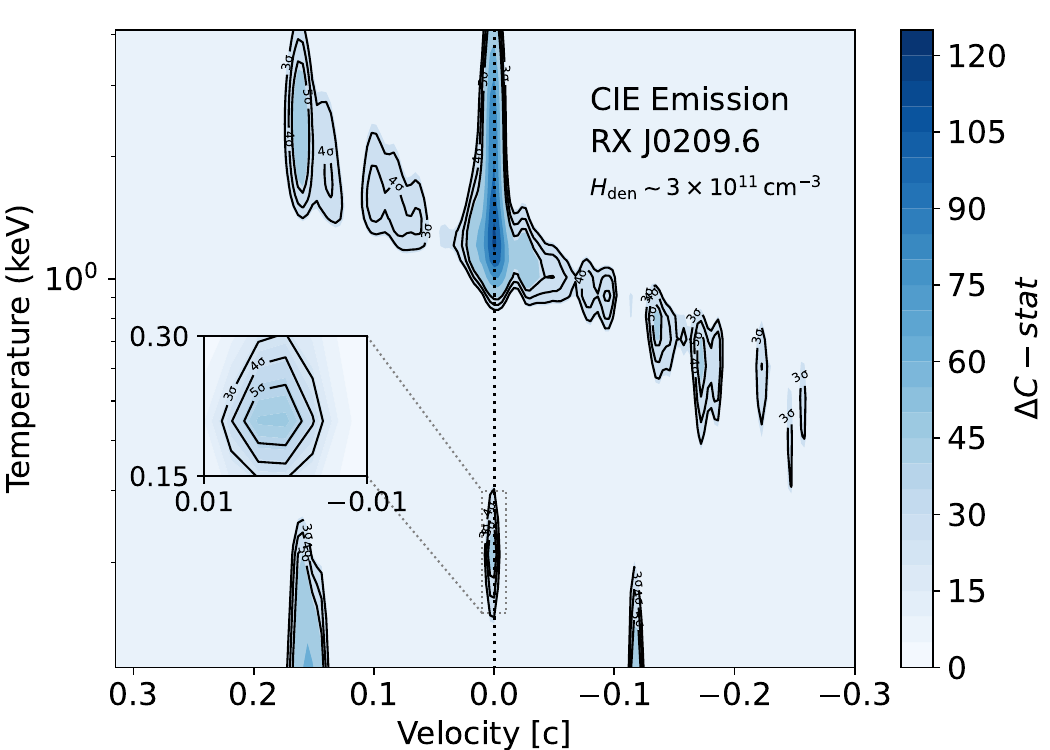}
    \includegraphics[width=.45\linewidth]{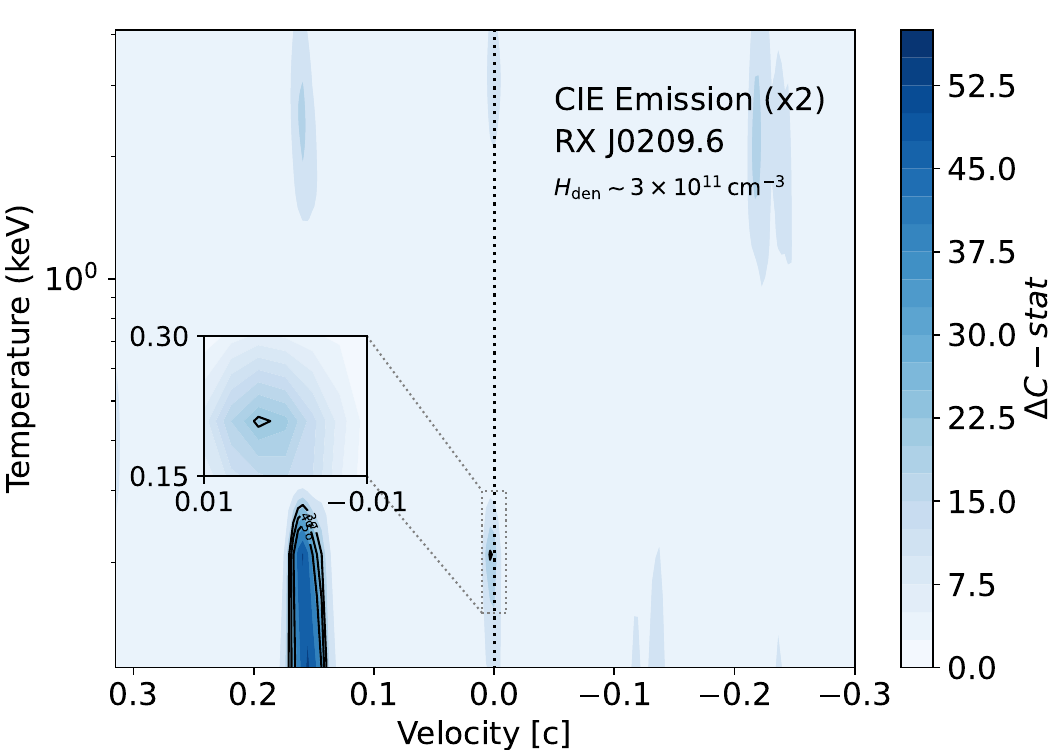}
    
    \caption{Grid search for collisionally ionised plasma in equilibrium, assuming a velocity width of $1000\ {\rm km\,s^{-1}}$, with one \texttt{cie} component (left) and two \texttt{cie} components (right) applied to the best-fitting continuum model. The contour lines indicate significance levels of 3, 4, and 5$\sigma$. The electron density of the emitting plasma was fixed at $3\times10^{11}\ {\rm cm^{-3}}$. The inset highlights the near-rest-frame solution associated with the modelling of the O {\scriptsize VII} triplet, corresponding to a plasma temperature of $kT \sim 0.2$ keV. The results obtained are consistent with the \texttt{cie} grids reported in the main text. }
    \label{fig:cie_emission_new_bin_hden}
\end{figure*}

\begin{figure*}[h!]
    \centering
\includegraphics[width=0.54\linewidth]{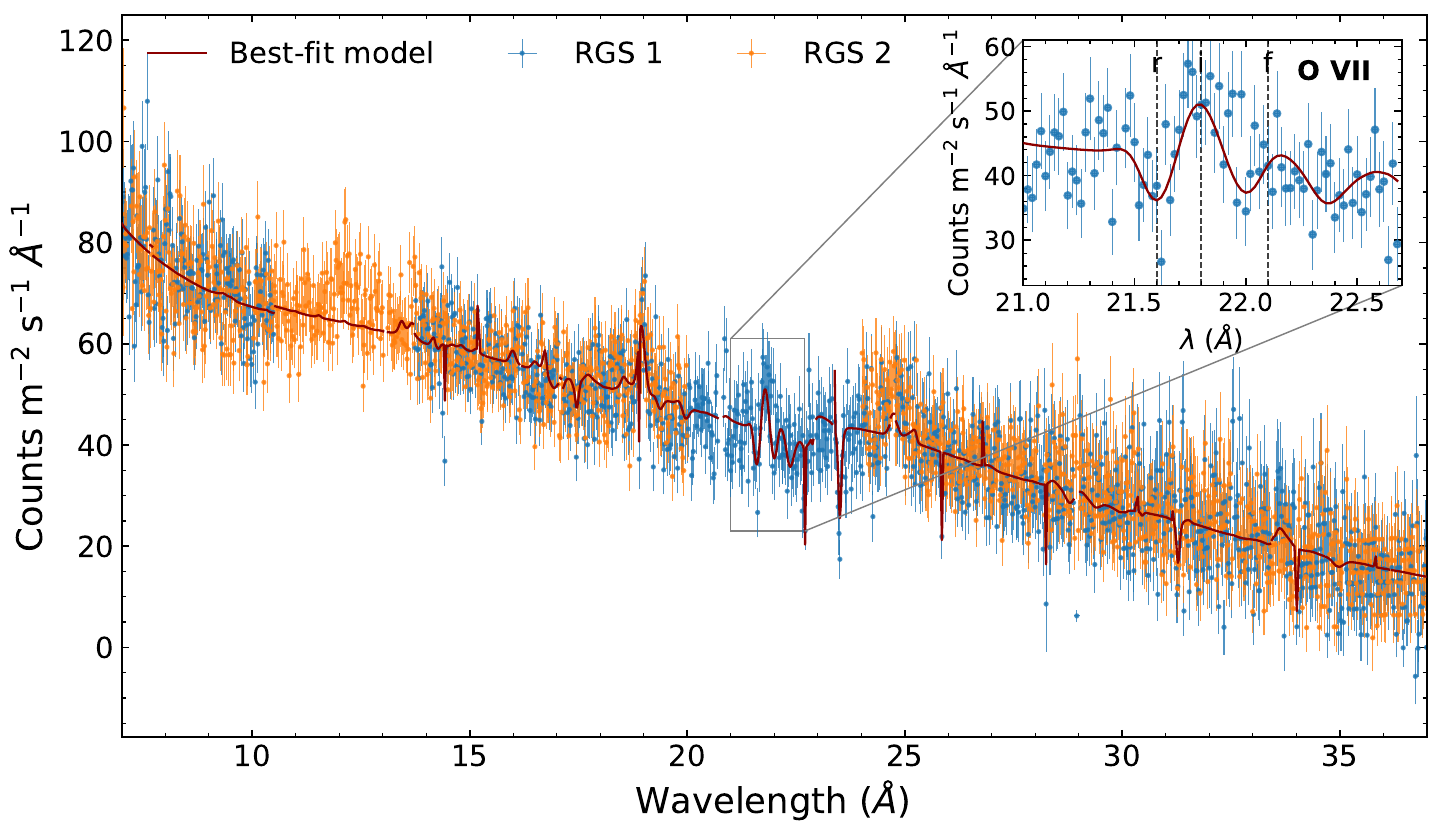}
  \includegraphics[width=0.45\linewidth]{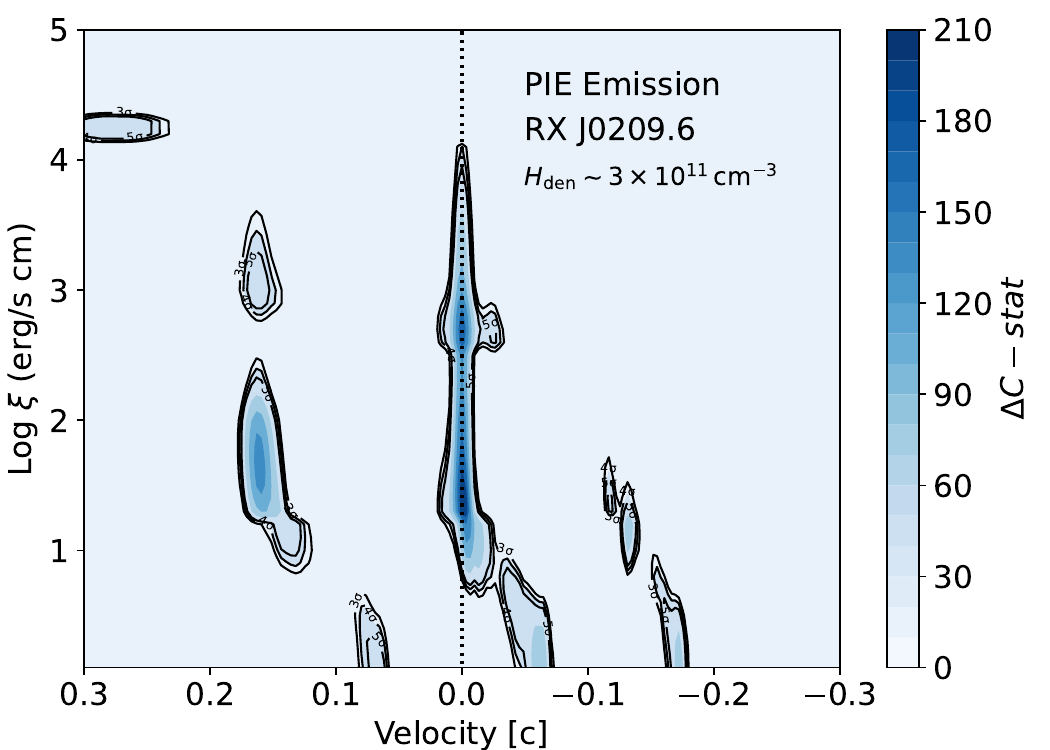}
    
   \caption{Best-fitting \texttt{hot*xabs*[continuum+pion]} model (left) and multidimensional scan grid of photoionised plasma model in emission (\texttt{pion}). The \texttt{xabs} (O {\scriptsize VII} resonance line absorber) component has a hydrogen column density of $N_{\mathrm{H}}=(2.34\pm0.26)\times10^{20}\ \mathrm{cm^{-2}}$, an ionisation parameter of $\rm log \ \xi  =0.88\pm0.05$, an RMS velocity broadening of $v_{\mathrm{rms}}=940\pm120\ \mathrm{km\,s^{-1}}$, and a systemic velocity of $v_{\mathrm{sys}}=120_{-140}^{+20}\ \mathrm{km\,s^{-1}}$. The \texttt{pion} component is characterised by an ionisation parameter of log $\xi=1.3 \pm 0.3$, an RMS velocity width of $v_{\mathrm{rms}}=1000\ \mathrm{km\,s^{-1}}$, and a hydrogen column density of $N_{\mathrm H}=(2.2\pm0.2)\times10^{20}\ \mathrm{cm^{-2}}$. The hydrogen number density is fixed at $n_{\mathrm H}=3\times10^{11}\ \mathrm{cm^{-3}}$, while the line-of-sight velocity is $v_{\mathrm{LOS}}=(-280\pm90)\ \mathrm{km\,s^{-1}}$. The contour lines indicate significance levels of 3, 4, and 5$\sigma$.} 
    \label{fig:1_PMF_absorber}
\end{figure*}

\begin{figure*}[h!]

    \centering
    \includegraphics[width=0.54\linewidth]{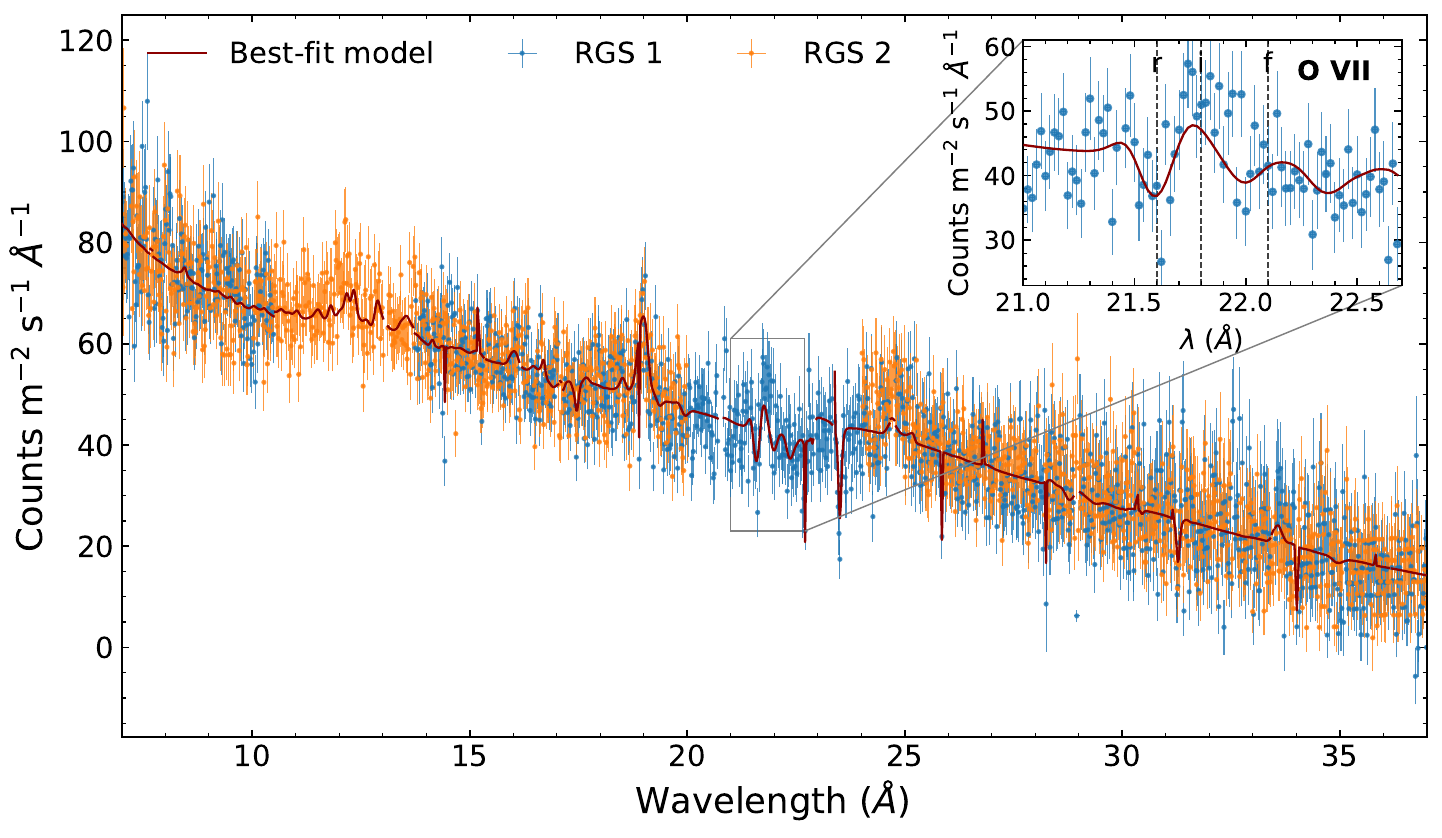}
  \includegraphics[width=0.45\linewidth]{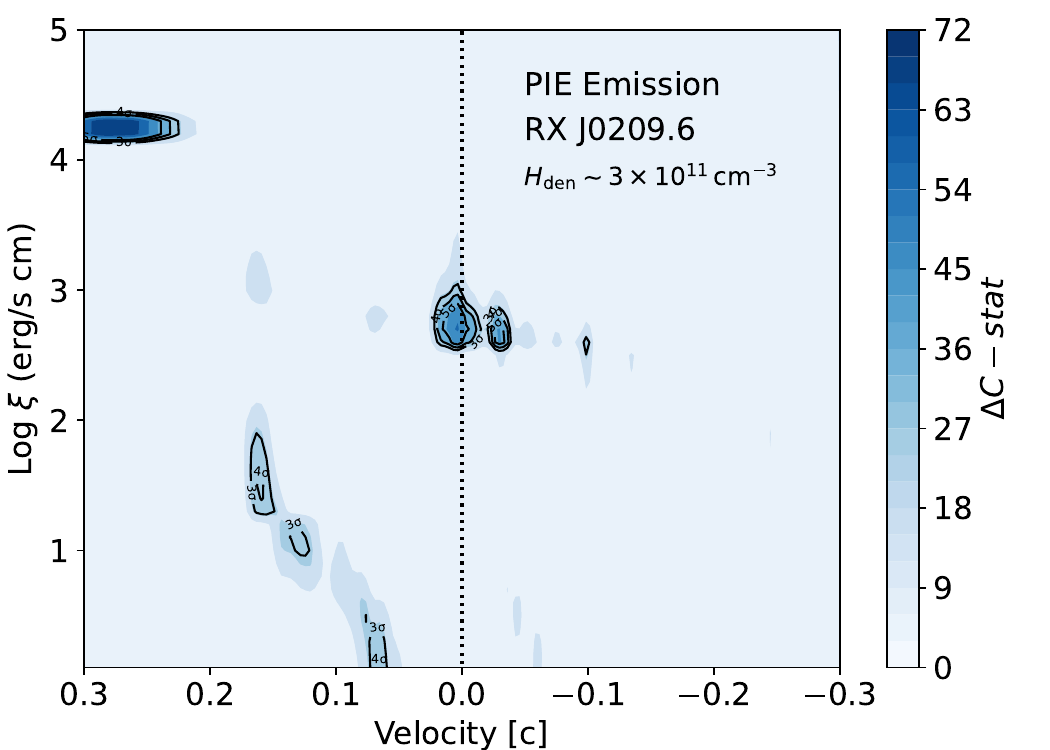}
    
   \caption{Best-fitting \texttt{hot*xabs*[continuum+2*pion]} model (left) and multidimensional scan grid of photoionised plasma model in emission (\texttt{pion}). The grid scan was produced by adding a second \texttt{pion} component and re-running the scan starting from the best-fit solutions identified in Fig. \ref{fig:1_PMF_absorber}. The \texttt{xabs} (O {\scriptsize VII} absorber) component has a hydrogen column density of $N_{\mathrm{H}}=(1.5\pm 0.3)\times10^{20}\ \mathrm{cm^{-2}}$, an ionisation parameter of $\log \xi =0.93_{-0.13}^{+0.10}$, an RMS velocity broadening of $v_{\mathrm{rms}}=900\pm200\ \mathrm{km\,s^{-1}}$, and a systemic velocity of $v_{\mathrm{sys}}=-320_{-260}^{+190}\ \mathrm{km\,s^{-1}}$.  The first \texttt{pion} component is characterised by an ionisation parameter of $\log\xi=1.3\pm0.3$, a hydrogen column density of $N_{\mathrm H}=(1.40\pm0.30)\times10^{20}\ \mathrm{cm^{-2}}$, and a line-of-sight velocity of $v_{\mathrm{LOS}}=(-870\pm250)\ \mathrm{km\,s^{-1}}$. The second \texttt{pion} component has an ionisation parameter of $\log\xi=2.7\pm0.1$, a hydrogen column density of $(1.32\pm0.15)\times10^{21}\ \mathrm{cm^{-2}}$, and a line-of-sight velocity of $v_{\mathrm{LOS}}=(700\pm300)\ \mathrm{km\,s^{-1}}$. Both components were calculated at a fixed hydrogen number density of $n_{\mathrm H}=3\times10^{11}\ \mathrm{cm^{-3}}$ and with an RMS velocity width of $v_{\mathrm{rms}}=1000\ \mathrm{km\,s^{-1}}$. The contour lines indicate significance levels of 3, 4, and 5$\sigma$. } 
    \label{fig:2_PMF_absorber}

\end{figure*}

\newpage

\section{Photoionised absorption grids after emission-line modelling }
\label{sec:absorber_ufo_with_rest_absorber}
\begin{figure*}[h!]

    \centering
    \includegraphics[width=.55\linewidth]{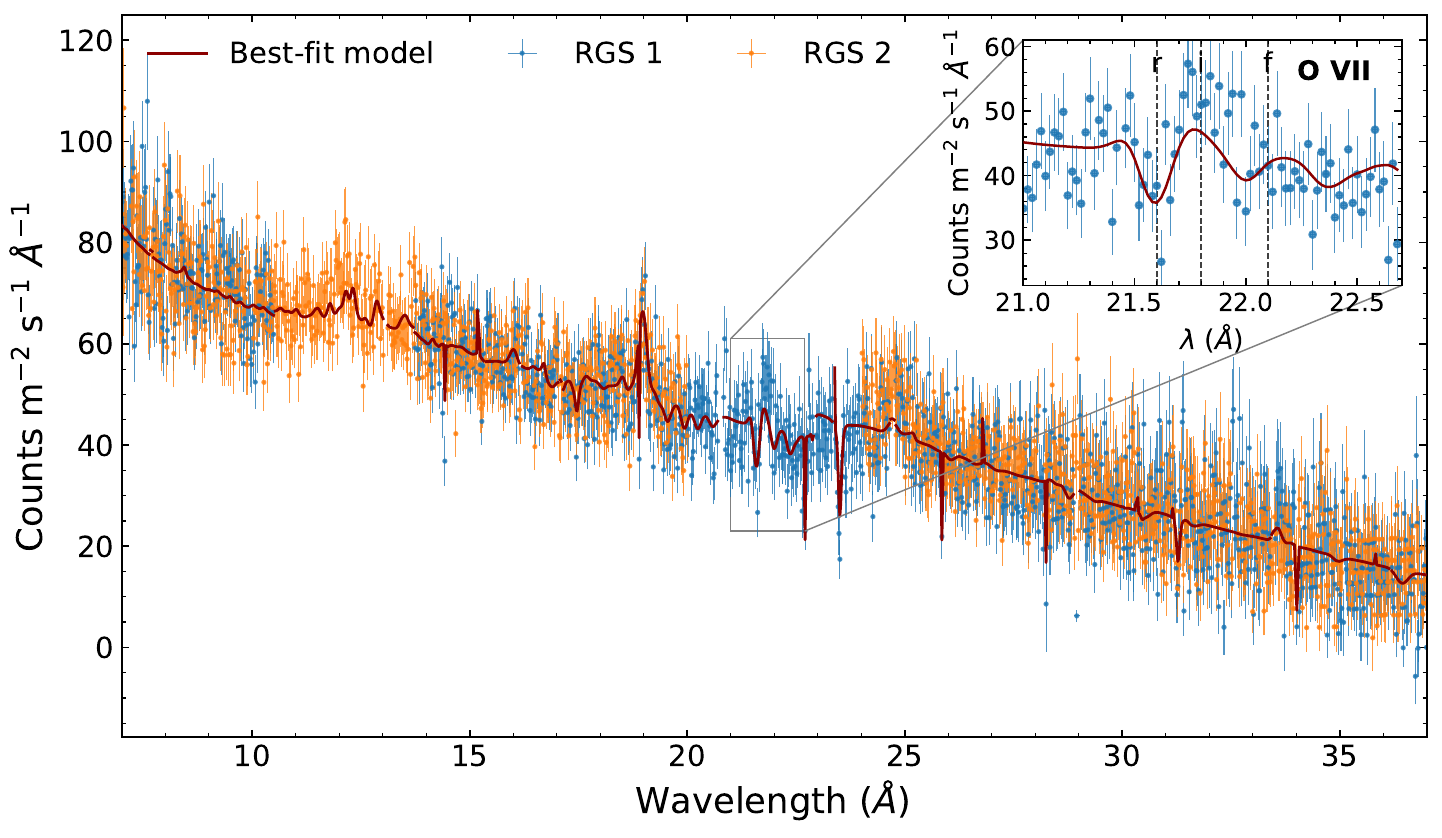}
    \includegraphics[width=.44\linewidth]{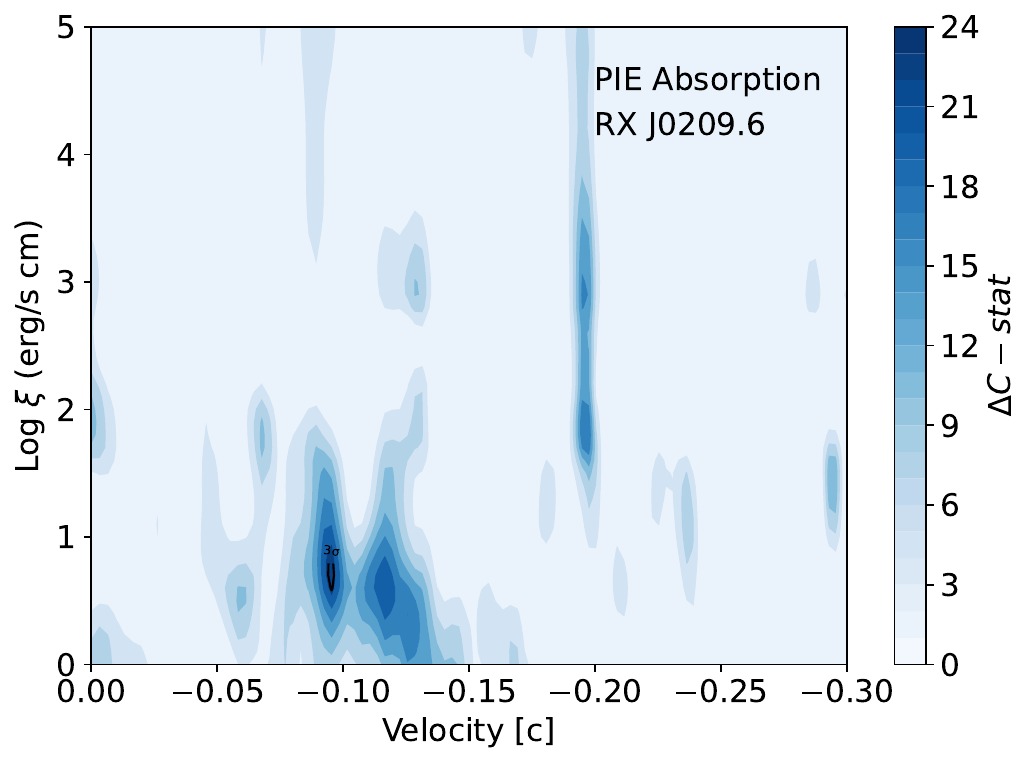}

   \caption{Spectral modelling (left) and grid search (right) for a relativistically blueshifted photoionised absorber. The two \texttt{pion} emission components were calculated
at a fixed hydrogen number density of $n_{\mathrm{H}}=3\times10^{11}\ \mathrm{cm^{-3}}$ and an RMS velocity dispersion of $1000\ \mathrm{km\,s^{-1}}$, as in the two-emitter configuration shown in Fig.~\ref{fig:2_PMF_absorber}. Left panel: best-fitting spectrum including the continuum, two \texttt{pion} emission components, a near-rest-frame \texttt{xabs} absorber accounting for absorption affecting the O {\scriptsize VII} resonance line, and a blueshifted \texttt{xabs} absorber corresponding to the solution at $v_{\mathrm{LOS}}\simeq-0.1c$ identified in the right-hand grid. The inset highlights the O {\scriptsize VII} triplet. Right panel: grid search for the blueshifted absorber, assuming photoionisation equilibrium. The baseline model includes the continuum, both \texttt{pion} emission components, and the near-rest-frame absorber. The contours mark the $3\sigma$ significance threshold.}
    \label{fig:xabs_emission_density_correct_1xabs}

\end{figure*}

\begin{figure*}[h!]

    \centering
\includegraphics[width=.54\linewidth]{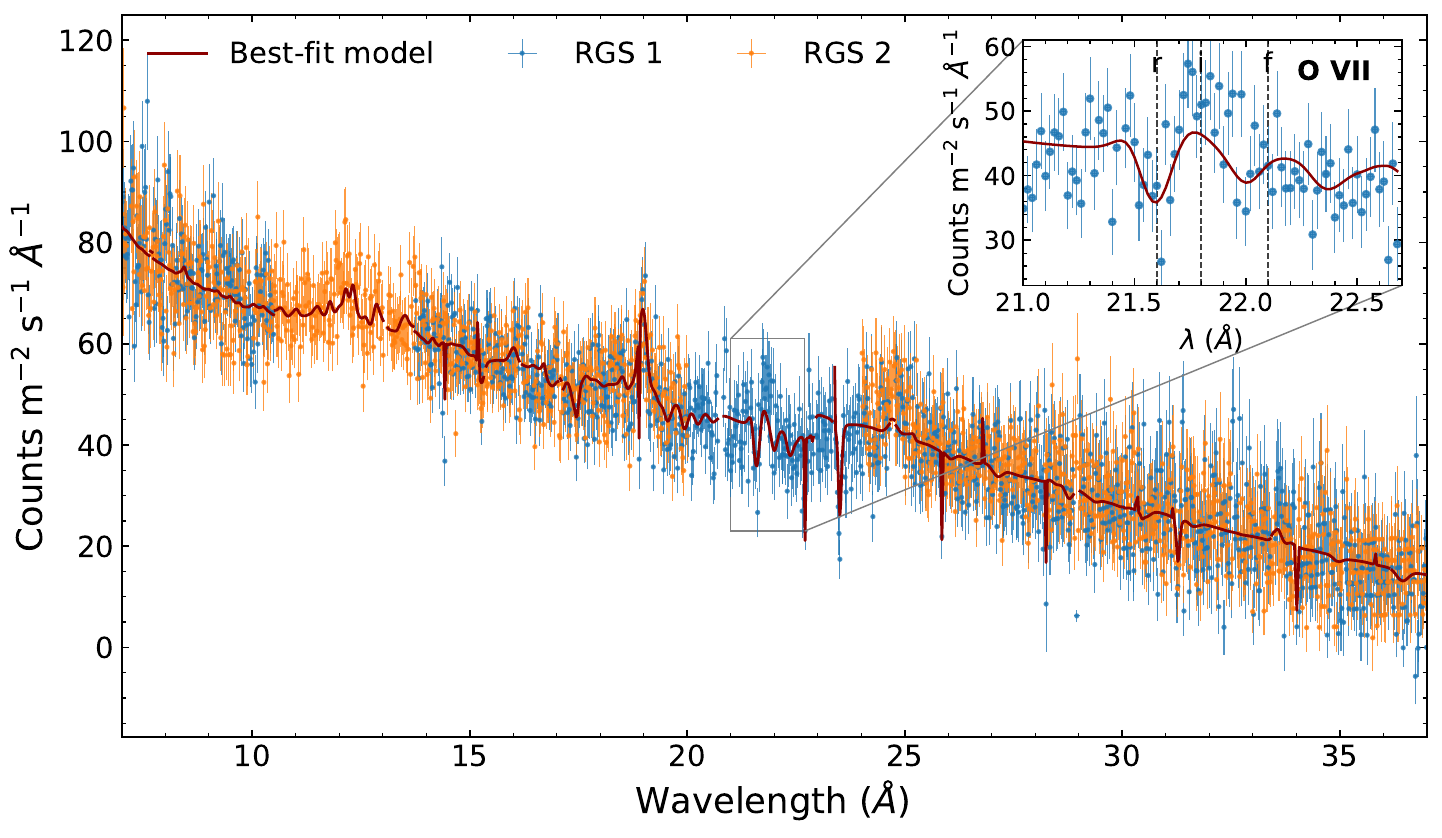}
    \includegraphics[width=.445\linewidth]{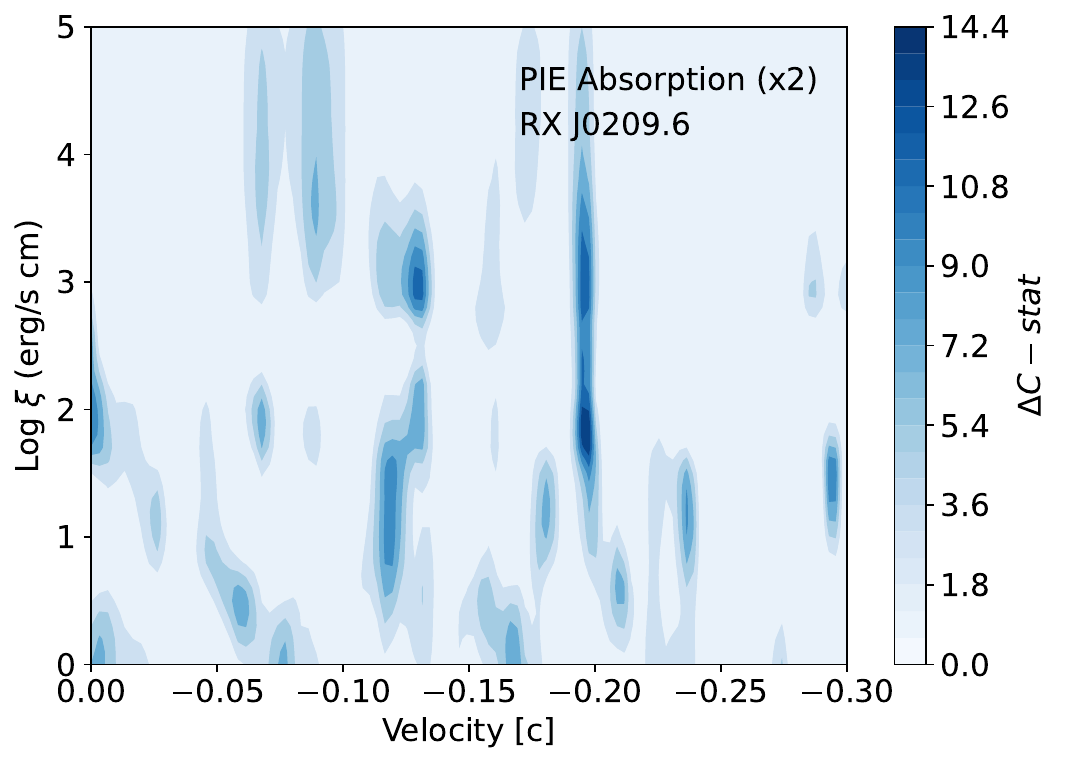}%

   \caption{Spectral modelling (left) and grid search (right) for a second relativistically blueshifted photoionised absorber. The two \texttt{pion} emission components were calculated at a fixed hydrogen number density of $n_{\mathrm{H}}=3\times10^{11}\ \mathrm{cm^{-3}}$ and an RMS velocity dispersion of $1000\ \mathrm{km\,s^{-1}}$, as in the two-emitter configuration shown in Fig.~\ref{fig:2_PMF_absorber}. Left panel: best-fitting spectrum including the continuum, two \texttt{pion} emission components, a near-rest-frame \texttt{xabs} absorber accounting for absorption affecting the O {\scriptsize VII} resonance line, and two relativistically blueshifted \texttt{xabs} absorbers. The second blueshifted absorber corresponds to the solution identified in the right-hand grid. The inset highlights the O {\scriptsize VII} triplet. Right panel: grid search for the second blueshifted absorber, assuming photoionisation equilibrium. The baseline model includes the continuum, both emission components, the near-rest-frame absorber, and the first blueshifted absorber at approximately $0.1c$, identified in Fig.~\ref{fig:xabs_emission_density_correct_1xabs}.}
    \label{fig:xabs_emission_density_correct_2xabs}

\end{figure*}

\end{appendix}

\end{document}